\documentclass{article}

\usepackage{microtype}
\usepackage{graphicx}
\usepackage{subcaption}
\usepackage{booktabs} 

\usepackage{hyperref}

\usepackage[preprint]{icml2026}

\usepackage{amsmath}
\usepackage{amssymb}
\usepackage{mathtools}
\usepackage{amsthm}
\usepackage{enumitem}
\usepackage{xspace}
\usepackage{xcolor}

\usepackage[capitalize,noabbrev]{cleveref}

\theoremstyle{plain}

\theoremstyle{definition}

\theoremstyle{remark}

\usepackage[textsize=tiny]{todonotes}

\newif\ifshowcomments
\showcommentstrue  

\newcommand{\para}[1]{\noindent\textbf{#1}}

\newcommand{\mechevalagent}{\texttt{MechEvalAgent}\xspace}

\icmltitlerunning{The Story is Not the Science: Execution-Grounded Evaluation of
Mechanistic Interpretability Research}

\begin{document}

\twocolumn[
  \icmltitle{The Story is Not the Science: \\ Execution-Grounded Evaluation of
Mechanistic Interpretability Research}



  \icmlsetsymbol{equal}{*}

  \begin{icmlauthorlist}
    \icmlauthor{Xiaoyan Bai}{uchi}
    \icmlauthor{Alexander Baumgartner}{equal,uchi}
    \icmlauthor{Haojia Sun}{equal,cmu}
    \icmlauthor{Ari Holtzman}{uchi}
    \icmlauthor{Chenhao Tan}{uchi}
  \end{icmlauthorlist}

  \icmlaffiliation{uchi}{University of Chicago}
  \icmlaffiliation{cmu}{Carnegie Mellon University}

  \icmlcorrespondingauthor{Xiaoyan Bai}{smallyan@uchicago.edu}

  \icmlkeywords{Machine Learning, ICML}

  \vskip 0.3in
]



\printAffiliationsAndNotice{}  

\begin{abstract} 
Reproducibility crises across sciences highlight the limitations of the paper-centric review system in assessing the rigor and reproducibility of research. AI agents that autonomously design and generate large volumes of research outputs exacerbate these challenges.
In this work, we address the growing challenges of scalability and rigor by flipping the dynamic and developing AI agents as research evaluators. 
We propose the first execution-grounded evaluation framework that verifies research beyond narrative review by examining code and data alongside the paper. 
We use mechanistic interpretability research as a testbed, build standardized research output, and develop \mechevalagent, an automated evaluation framework that assesses the \emph{coherence} of the experimental process, the \emph{reproducibility} of results, and the \emph{generalizability} of findings. We show that our framework achieves above 80\% agreement with human judges, identifies substantial methodological problems, and surfaces 51 additional issues that human reviewers miss.
Our work demonstrates the potential of AI agents to transform research evaluation and pave the way for rigorous scientific practices.\footnote{Code available: \url{https://github.com/ChicagoHAI/MechEvalAgent/}} 
\end{abstract}

\section{Introduction}

\begin{figure*}[t]
    \centering
    \includegraphics[width=0.99\textwidth]{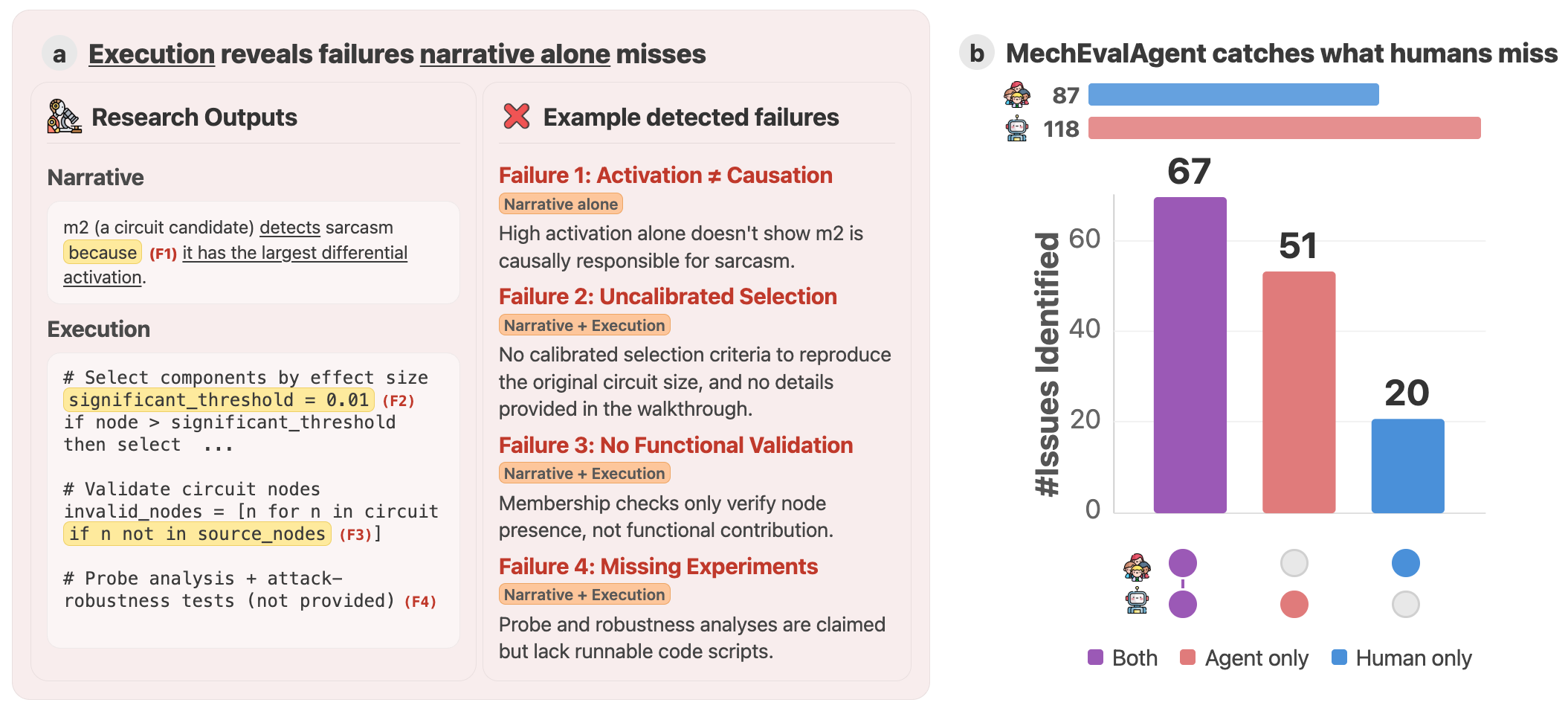}
    \caption{(a) Execution-grouned evaluation uncovers failures that narrative-alone review misses. In this example, 
    Failures 2, 3, and 4 require execution beyond narrative review. (b) As a highlight of our results, we find that \mechevalagent surfaces 51 additional issues that human reviewers overlooked.}
    \label{fig:venn}
\end{figure*}

Peer review has long treated the paper narrative as the primary object of evaluation~\cite{kelly2014peer}.
While this may suffice for theoretical work, it poses challenges for empirical research, which cannot be fully assessed without executable artifacts such as code and data.
Reproducibility crises across scientific fields have exposed the limitations of narrative-alone review~\cite{open2015estimating,desai2025reproducibility,RetractionWatch2024Psychology}.
In one case, manipulated behavioral data 
resulted in a high-profile faculty termination at a leading research university~\cite{DataColada2024GinoReport}.
In the context of AI research, even when code is shared, reviewers often lack the time or resources to run it, leading to a reliance on narrative descriptions that may not accurately reflect the actual implementation or results.

Agentic workflows add more challenges as they are now widely deployed in research tasks, including code assistance~\cite{anthropic2025claudecode,openai2025codex,novikov2025alphaevolve}, ideation~\cite{zhou2024hypothesis,baek2025researchagent}, and end-to-end research~\cite{schmidgall2025agent,ifargan2025autonomous}. They autonomously generate large volumes of research outputs and can accelerate publication timelines, 
contributing to even more submissions in an already strained system. 
Notably, the number of papers submitted to NeurIPS increased from 1,420 submissions to 21,575 submissions from 2013 to 2025, a 15-fold increase in submission~\cite{papercopilot_neurips_statistics,neurips2025_review_reflections,desai2025reproducibility}.
AI involvement also exacerbates the challenges of narrative-alone evaluation, as implicit hallucinations are hard to detect~\cite{haibe2020transparency,beam2020challenges}.

Recent efforts have begun to use AI in the evaluation process to address this challenge, including systems that provide paper feedback~\cite{PaperReviewAI_Stanford_2025} and broader uses of LLMs or agents in peer review~\cite{lee2025role,liang2024can,yu2024researchtown}. These systems can already detect important issues such as hallucinated references~\cite{GPTZero_NeurIPS_Hallucination_Check_2026}, and studies show that AI generated reviews overlap well with human reviews~\cite{liang2024can}. However, most approaches 
examine
the coherence and consistency of the story in the paper and ignore all the executable resources.
In other words, the focus is on how appealing the story is rather than whether the story is true.\footnote{Narrative-alone evaluation is even more concerning in AI review, where authors have inserted prompts in papers to raise the scores of AI judges~\cite{nature_ai_peer_review_2025}.}

In this work, we seek to verify the science itself, not just the story told about it.
Motivated by this view, we propose a general evaluation framework that combines narrative and execution to reveal errors that narrative-alone review misses, as shown in Figure~\ref{fig:venn}, demonstrating how agents can support rather than undermine scientific review.

We propose the first execution-grounded evaluation framework that standardizes research outputs by bundling execution resources with narrative. This enables systematic assessment of \emph{coherence} of the experiment process, \emph{reproducibility} of the results, and \emph{generalizability} of the findings beyond what the final paper alone can provide.

We build \mechevalagent, an automated evaluation agent that implements our pipeline.  
We focus on mechanistic interpretability as a testbed because its claims are often expressed in terms of concrete mechanisms and interventions that can be directly tested by rerunning experiments and applying the proposed analyses to new models or inputs. The field also follows relatively standardized methodological patterns, which makes failures in execution and evaluation easier to isolate. Finally, generalizability is a central open question in mechanistic interpretability, making it a natural setting to test whether evaluation can go beyond narrative.
While we instantiate our framework in mechanistic interpretability, 
the framework itself is not domain-specific and can be adapted to other areas of scientific research. 

We evaluate \mechevalagent on 30 research outputs in mechanistic interpretability. \mechevalagent achieves above 80\% agreement with human experts. As shown in Figure~\ref{fig:venn}b, it captures most failures identified by humans (67 of 87) and surfaces 51 \textit{additional} issues that humans missed. Many of these issues illustrated in Figure~\ref{fig:venn}a require execution to detect, such as missing data selection criteria that only emerge during reproduction (Failure 2), validation code that checks list membership rather than task validity (Failure 3), or missing files that prevent reproducing key results (Failure 4). Removing code access or execution substantially reduces agreement to about 45\% on average, and \mechevalagent evaluates faster than humans, who take 2.2 hours per task on average, aligned with research in other disciplines~\cite{abogunrin2025much}. These results highlight the value of execution-grounded evaluation.

To summarize, our main contributions are as follows:

\begin{itemize}[topsep=0pt,itemsep=-2pt,leftmargin=*]
    \item We develop the first execution-grounded evaluation framework for research outputs that goes beyond narrative-alone review by standardizing research outputs and building systematic evaluation suites.
    \item We build \mechevalagent to instantiate this framework in mechanistic interpretability and evaluate 30 research outputs from both humans and AI research agents.
    \item Our evaluation results align with human experts and surface 51 issues overlooked by human reviewers, demonstrating the importance of combining narrative and execution-grounded assessments.
\end{itemize}

\section{Method} \label{sec:method}

\mechevalagent expects research outputs to include both narrative and execution resources. We standardize research outputs and introduce an evaluation framework based on these outputs to enable execution-grounded assessment.

\para{Unified research outputs.}
We propose a unified standard for research outputs. Research outputs are expected to include two components:
\begin{itemize}[topsep=0pt,itemsep=0pt, leftmargin=*]
    \item \textbf{Narrative}: For human-written papers, we extract the 
    the \emph{plan} and the \emph{report}. For agent-generated outputs, we 
    expect the research trace beyond a final paper, including \emph{human prompts}, for instance. The plan specifies the goal, hypothesis, constraints, and intended methodology. Comparing it against the executed code enables checks of plan-implementation consistency. This helps detect goal drift and cases where an intended methodology is quietly replaced by a different method that yields favorable results. The report summarizes goals, methods, results, and conclusions, links claims to evidence, and can briefly discuss future directions and implications.
    \item \textbf{Execution Resources}: We expect the \emph{implementation}, including code and data, together with a \emph{walkthrough} of it. 
    This enables grounded checks of runnability and correctness, and surfaces failures that are invisible in narrative-alone review, such as broken environments, API mismatches, or incorrect metric computations. The walkthrough also allows us to assess whether sufficient information is provided for reproduction.
\end{itemize}
These resources make it possible to verify not only \emph{what} an agent claims, but also \emph{how} the claim was produced. For example, a report may state that an intervention improves a metric, but execution can reveal that the code does not run, implements a different computation than described, or fails to reproduce the reported numbers.

\para{Our proposed evaluation framework.}
\begin{figure}[t]
    \centering
    \includegraphics[width=0.47\textwidth]{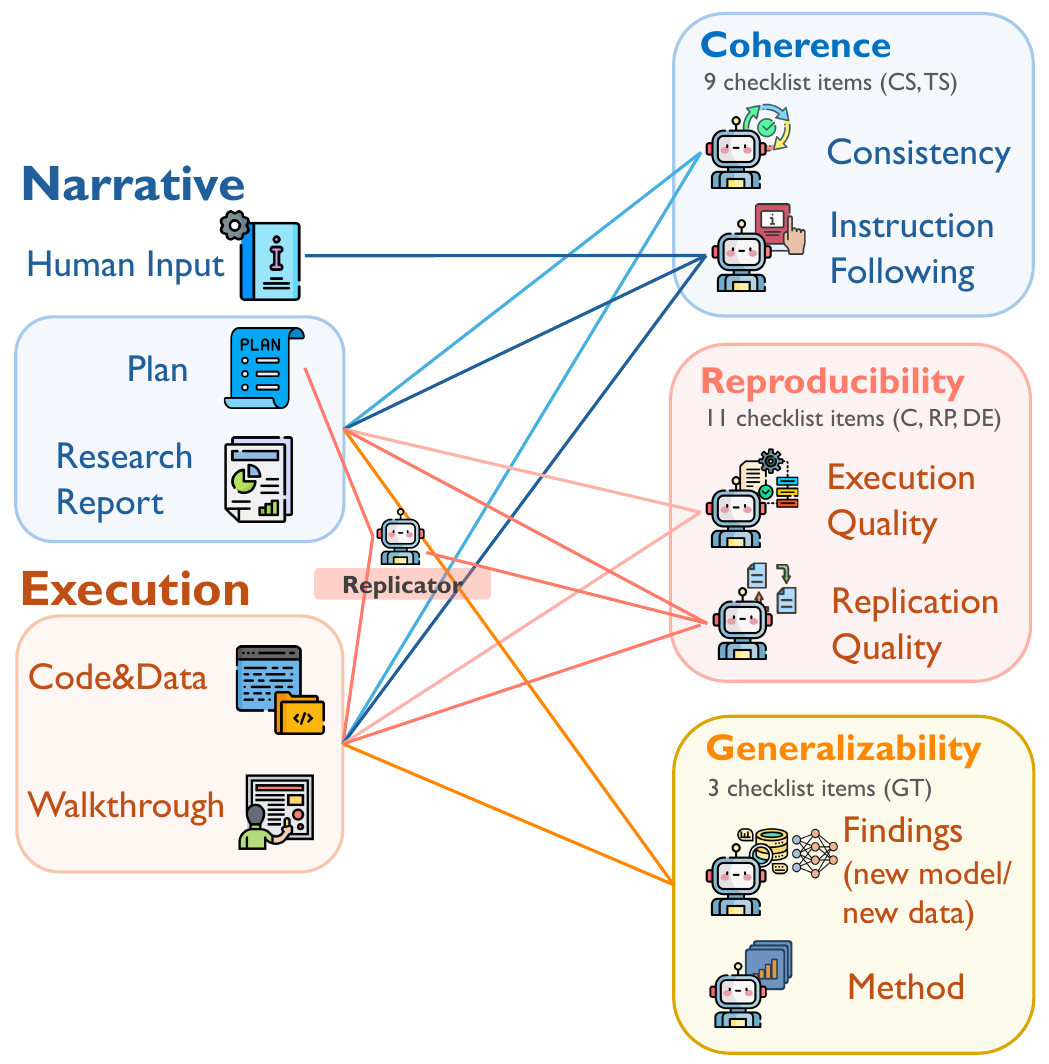}
    \caption{Overview of the \mechevalagent framework. Research outputs are evaluated on coherence, reproducibility, and generalization, with each sub-dimension handled by an agent that takes in the relevant inputs.}
    \label{fig:pipeline}
\end{figure}
Our evaluation framework formalizes the review process into explicit checks. When evaluating research outputs, a reviewer typically wants to know three things: (i) whether the claims are consistent with the provided evidence, (ii) whether the results can be reproduced, and (iii) whether the findings generalize beyond the original setting. 
As illustrated in Figure~\ref{fig:pipeline}, our \mechevalagent operationalizes these questions in the following three components, with more detailed description in Appendix~\ref{sec:app_checklist}.

\para{Coherence.} 
Coherence evaluates on \emph{consistency} and \emph {instruction following}. It asks whether the research outputs are internally consistent and aligned with its stated goals, including whether the implementation and reported results support the claims and whether the artifact follows the intended research objective.

\para{Reproducibility.}
Reproducibility evaluates on \emph{execution quality} and \emph{replication quality}. It asks whether the data is presented, the code is runnable and correctly implements the described computations, and whether an independent replication run can reproduce the reported results. We forbid the access to report to avoid hallucination. As shown in Figure~\ref{fig:pipeline}, we have another agent to evaluate on the replication quality including check whether there is external reference and hallucination in replication and whether the result and the conclusion matches with the original ones.

\para{Generalizability.}
In this part, an agent reads the research repository and checks whether the reported findings generalize beyond the original setting, including to a new model, new data instances, or related tasks. This evaluation process measures whether one can learn something from the research by testing the predictions based on the findings in a new situation, considered an important component of generalizability~\cite{kim2025llmspursueagenticinterpretability}.
This evaluation is well-suited to mechanistic interpretability, where research often identifies mechanisms in a specific model and task while requiring generalizable insights across architectures or contexts. Therefore, whether the agent is able to do this serves as an indirect measure of whether the original research captured generalizable insight rather than task- or model-specific patterns.

In practice, we evaluate each component through structured binary checklists. 
The advantage is that checklists reduce subjectivity and enable consistent aggregation of evaluation results across agents and tasks and enables comparisons with human experts.\footnote{Anecdotally, we find that \mechevalagent can identify even more issues when the checklist is not provided, as it can explore more aspects of the research output. However, for systematic evaluation and comparison, we focus on checklist-based evaluation in this work.}
Figure~\ref{fig:pipeline} summarizes the checklist structure and overall workflow. Each sub-dimension is implemented by a dedicated agent that evaluates multiple items in the checklist. For example, our checklists cover statistical significance (\texttt{CS5}), effect size (\texttt{CS3}), and justification (\texttt{CS4}) for \emph{coherence}, code runnability (\texttt{C1}) for \emph{reproducibility}, and generalization to new models and data instances (\texttt{GT1, GT2}) for \emph{generalizability}.\footnote{Checklist item of statistical significance is captured by top \href{https://neurips.cc/public/guides/PaperChecklist}{ML conference}, as well as reproducibility of the main results and broader impact of the research. }  The full list of checklist items is included in Table~\ref{tab:checklist}, and additional implementation details and example outputs are included in Appendix~\ref{app:evaluator_details}.

\para{Building \mechevalagent.}
We develop \mechevalagent on Claude Code, with execution and logging support provided by Scribe~\cite{goodfire2025scribe}. Scribe integrates code generation with Jupyter notebooks, making execution traces and errors explicit. This enables reliable execution checking and qualitative analysis. Specifically, we reorganize both narrative and execution resources, including input prompts to the research agent, plans, code implementations, relevant data, walkthroughs, and final reports. The structure for human-written repositories differs slightly and is explained in Section~\ref{sec:exp_setup}. We introduce specialized agents responsible for evaluating different metrics as discussed above. During early iterations, we identified a risk of \mechevalagent modifying source files. To prevent this, we enforce strict file access restrictions and integrate with GitHub for version control monitoring to ensure no unintended modifications occur. To guard against hallucination in reproduction, we route replication results through a separate verification agent that checks result fidelity. With this modular, execution-grounded design, \mechevalagent can be extended to other domains beyond mechanistic interpretability. Figure~\ref{fig:pipeline} illustrates the detailed input structure of \mechevalagent.\footnote{Please refer to supplementary materials for detailed input templates.}

\section{Experiment Setup}\label{sec:exp_setup}
To cover a diverse range of research scenarios, we evaluate our framework and \mechevalagent on three types of research outputs, with ten examples each. The selected tasks are shown in Table~\ref{tab:tasklist}:

\begin{itemize}[leftmargin=*,topsep=-2pt,itemsep=-2pt]
    \item \textbf{Replication of Research.}
The agent is given a concrete research goal and hypothesis from an established paper and must generate a plan and implement experiments that directly test it. This setting tests instruction following in a realistic workflow where a human delegates a well-defined experiment. It is also prone to hallucinated reasoning, since the agent may rely on memorized conclusions rather than grounded execution. Our \mechevalagent surfaces such failures by checking whether the implementation and results actually support the stated hypothesis.
    \item \textbf{Open-ended Research Questions.}
Agents are given open-ended questions with only high-level goals (e.g., identifying sarcasm-related circuits), without a known ground truth. This setting reflects exploratory research, where agents must form their own hypotheses and outcomes can be either positive or negative. Coherence is especially important at this stage, since an agent should be able to report null or negative results rather than defaulting to overly optimistic conclusions.
    \item \textbf{Human-Written Repositories.}
Our evaluation is not limited to agent-generated research. Human-written repositories can exhibit similar issues, such as overclaiming or poorly reproducible code. Since these repositories often lack explicit plan files, we extract the goal, claims, and methodology from the accompanying paper to apply our pipeline in a unified manner. In this setting, we evaluate all aspects of the pipeline except instruction following, which is not applicable. 
\end{itemize}

Research agents naturally produce richer outputs than traditional artifacts, including plans, intermediate logs, and human prompts. We fully utilize and standardize these into a unified format for systematic checklist evaluation, enabling checks on plan-implementation consistency and verification that conclusions follow from execution. Our research agent is built on Scribe, which processes our instruction prompts and structures the research output according to our standardized format. We include an example of the detailed prompt in Appendix~\ref{app:sub_experiment_setup}.

\para{Evaluating our evaluation framework.}
To evaluate our evaluation framework and \mechevalagent, we first measure \textbf{agreement} between human evaluators and \mechevalagent. Human evaluators are asked to use the exact same checklists and instructions as the agents to assess the research \textit{independently}, including reproducing the repository and testing generalizability. We measure the agreement between human and agent judgments. This evaluates whether our agents produce evaluations that are consistent with human reasoning. We also examine evaluator \textbf{efficiency} by measuring the time humans spent on evaluation. 
In addition, we evaluate the \textbf{rated quality} of the agent's evaluation output. Human experts assess whether the issues reported by our agents are correct, meaningful, and complete by scoring their assessment with the agents from 1  to 5 (strongly disagree - strongly agree with the agent's judgement). Due to limited resources, we have three human authors as annotators.

We run each automated evaluation three times. We use AND  logic for PASS (equivalent to OR logic for FAIL), where a task is marked as PASS only when all runs return PASS.
We additionally analyze majority vote and cross-run stability in Appendix~\ref{sec:appendix_results}.
Due to limited resources, we asked three human experts to evaluate 30 research generations, with each generation evaluated by exactly one expert. Each expert also evaluated the quality of the automated evaluation output for a single evaluation run.

\para{Ablation study.}
To isolate the value of execution, we compare \mechevalagent to two ablated evaluators. The \emph{Doc-Only} evaluator reads only the final report and cannot inspect or run code, approximating limited paper-only review. The \emph{No-Execution} evaluator sees the full repository but cannot execute it, so reproducibility and generalizability are only judged from written evidence rather than executions. More details of the setting are shown in Appendix~\ref{app:sub_experiment_setup}.

\section{Results}\label{sec:results}

Our results show that execution failures and narrative weaknesses are common across tasks, undermining reliability. \mechevalagent aligns with human judgments while surfacing additional issues that humans miss. Ablations further confirm the value of execution-grounded evaluation.

\subsection{An Overview of Key Findings}
\begin{figure}[t]
    \centering
        \includegraphics[width=0.40 \textwidth]{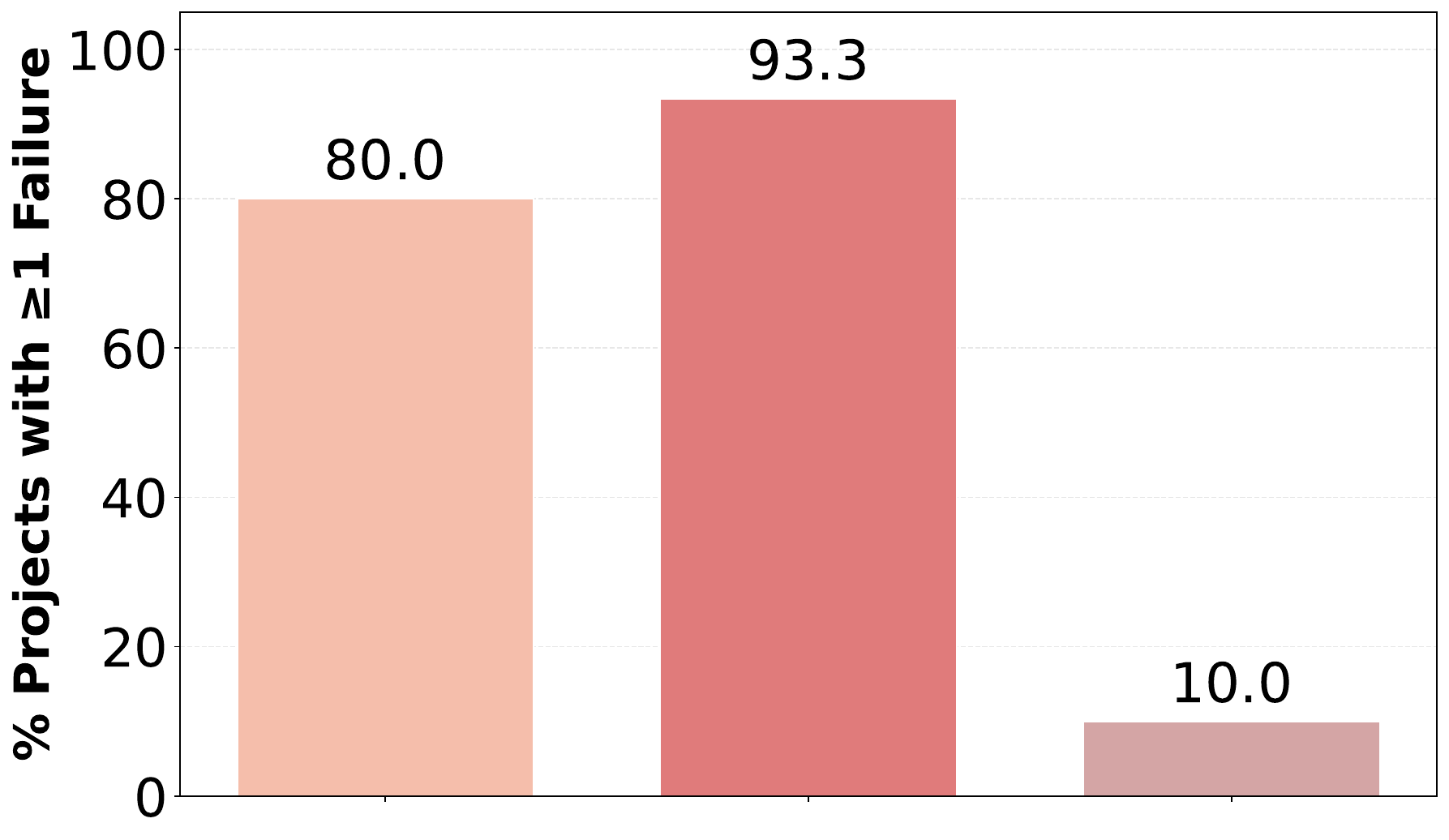}
    \includegraphics[width=0.5\textwidth]{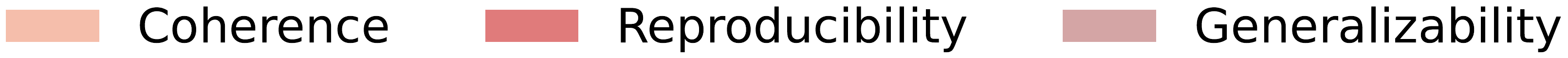}
    \caption{Percentage of projects with at least one failure per dimension. Over 90\% of tasks fail in reproducibility, and 80\% fail in coherence.}
    \label{fig:and_result}
\end{figure}

\para{Execution and narrative issues are prevalent in research outputs (Figure~\ref{fig:and_result}).} 
Over 90\% of the tasks have at least one failure in reproducibility, largely driven by execution errors, and 80\% of the tasks fail in coherence, mainly due to lack of consistency, as shown in Figure~\ref{fig:and_result} and Figure~\ref{fig:failure_all_metrics}. Execution failures often stem from model loading, shape calculations, and environment setup, and they appear in both agent-generated and human-written repositories. Our checklists also surface narrative weaknesses in evidential support, such as unreliable effect sizes, weak statistical significance, or conclusions based on insufficient results.

\begin{figure}[t]
    \centering
    \includegraphics[width=0.4\textwidth]{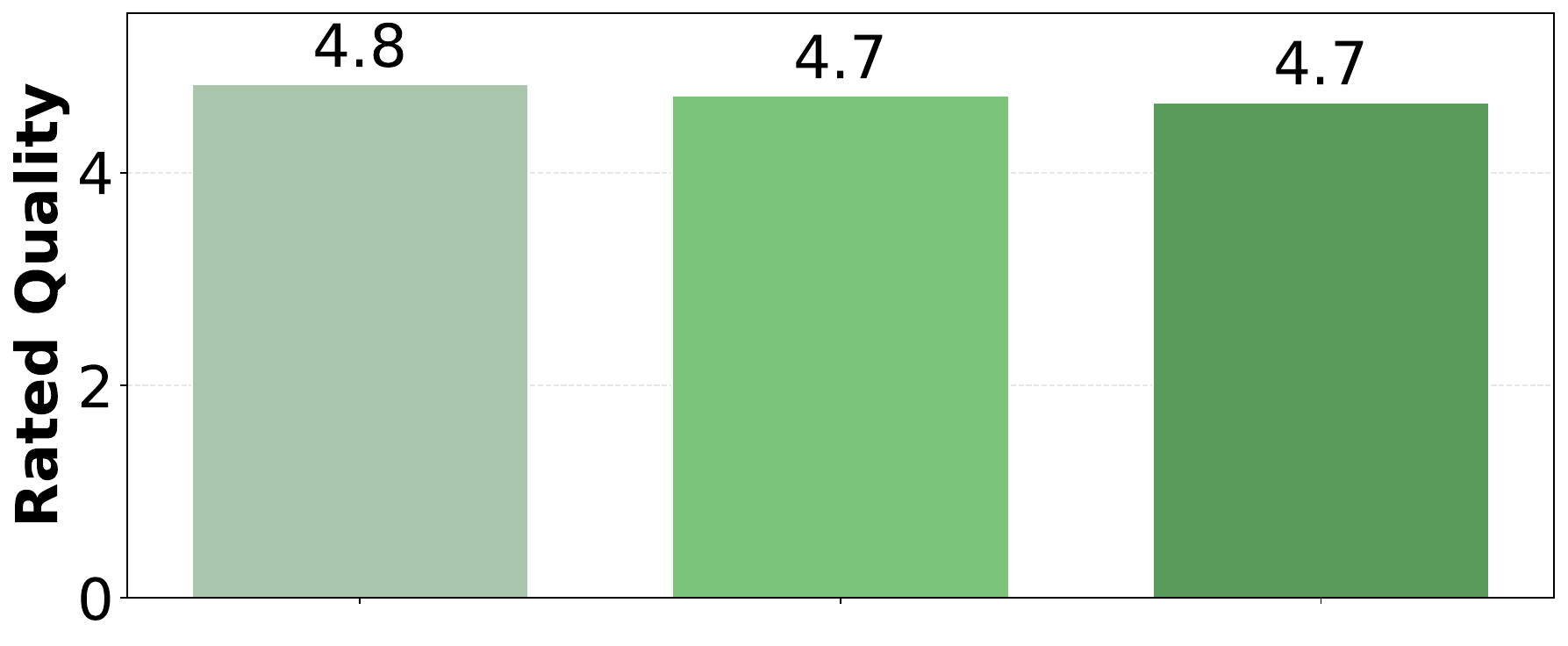}
    \includegraphics[width=0.47\textwidth]{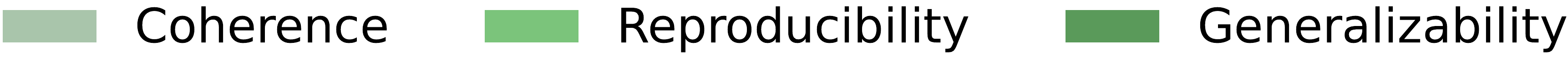}
    \caption{Human-rated quality on \mechevalagent evaluations (1-5 Likert scale, 1 = Strongly Disagree, 5 = Strongly Agree). All dimensions show ratings above 4.7, indicating high quality of agent assessments.}
    \label{fig:human_rating}
\end{figure}

\begin{figure}[t]
    \centering
    \includegraphics[width=0.4\textwidth]{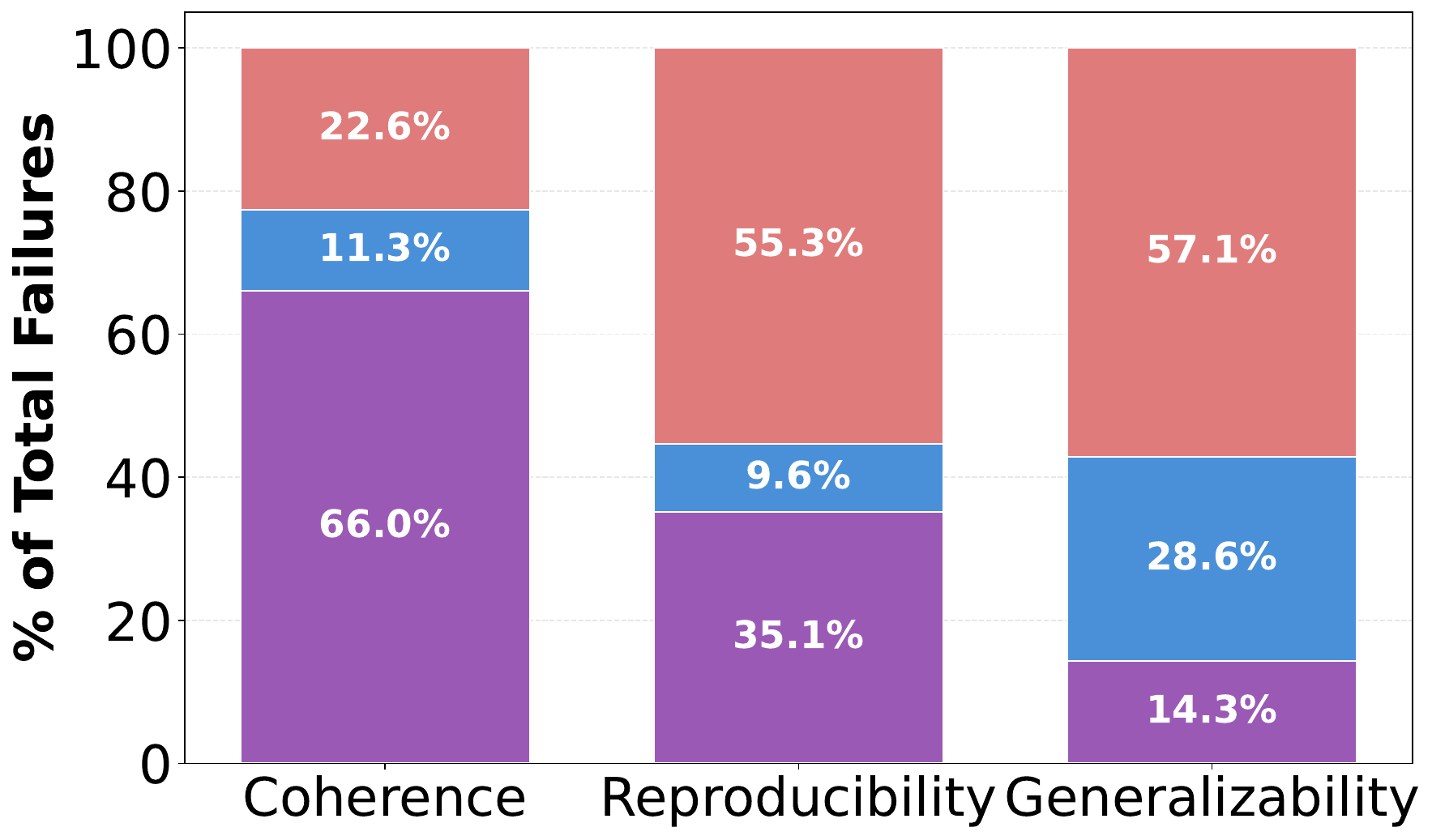}
    \includegraphics[width=0.45\textwidth]{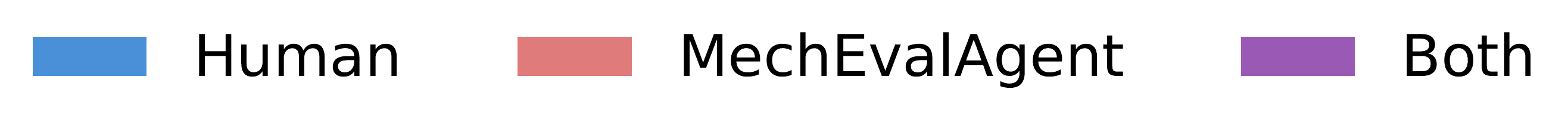}
    \caption{Failure breakdown comparing human-identified and agent-identified issues. \mechevalagent surfaces more unique issues in all three dimensions.}
    \label{fig:failure_breakdown}
\end{figure}

\begin{figure}[t]
    \centering
    \includegraphics[width=0.4\textwidth]{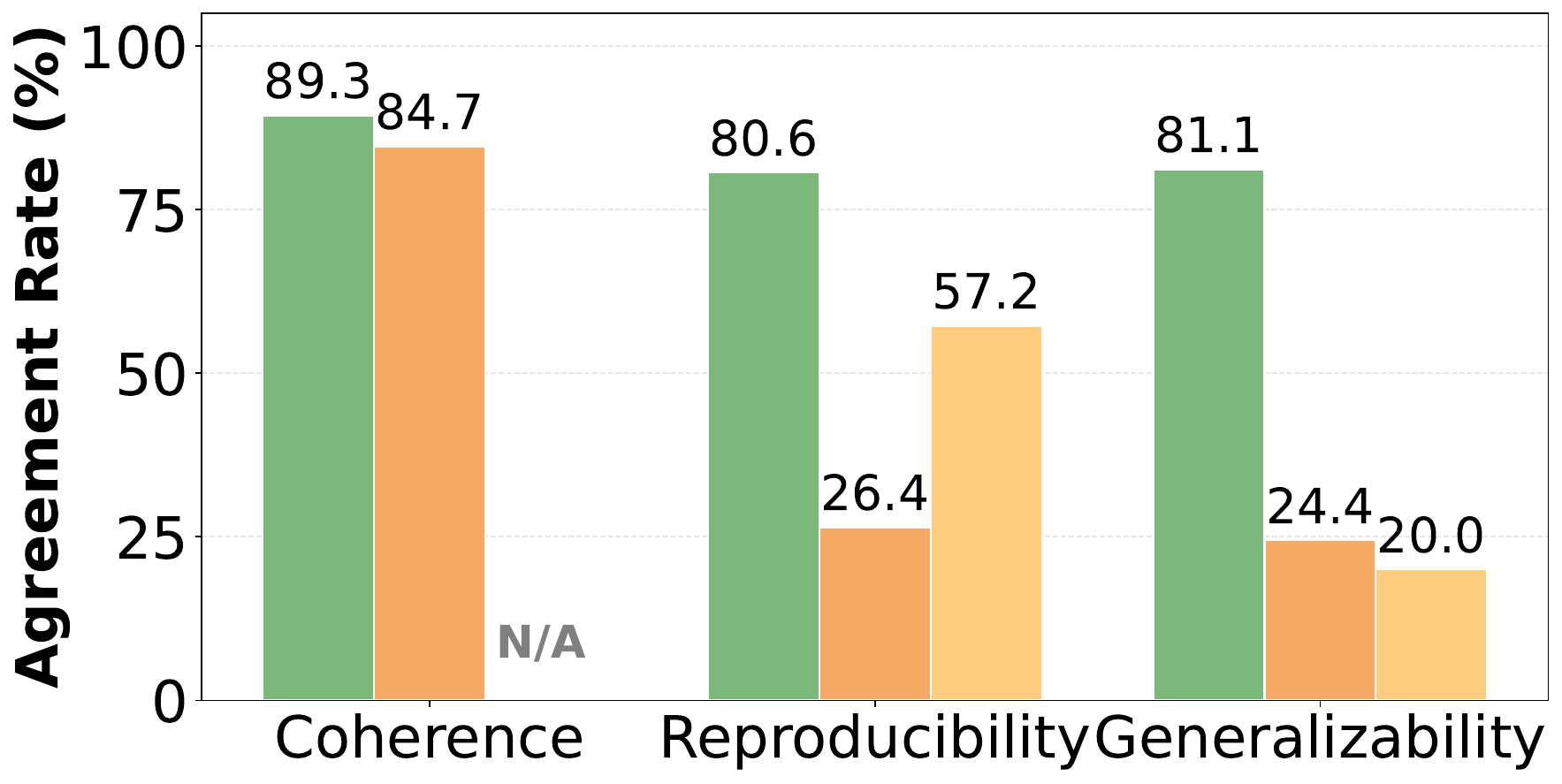}
    \includegraphics[width=0.47\textwidth]{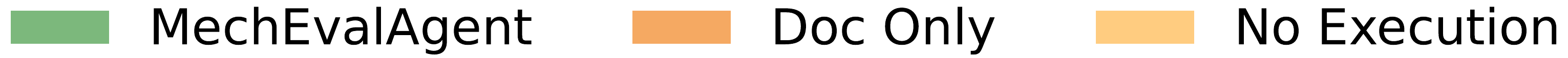}
    
    \caption{Agreement with human experts across the full \mechevalagent  pipeline and two ablated variants (Doc-Only and No-Execution). Full \mechevalagent pipeline shows high agreement across all dimensions. Both ablations perform substantially worse than the full pipeline.}
    \label{fig:ablation}
\end{figure}

\para{\mechevalagent aligns with human judgments, while execution helps surface additional issues beyond narrative.}
As shown in Figure~\ref{fig:human_rating}, 
rated quality of the evaluation outputs by \mechevalagent is high, with scores above 4.7 out of 5 across all dimensions.
The agreement is above 80\% across all dimensions (Figure~\ref{fig:ablation}).
Figure~\ref{fig:venn} shows that \mechevalagent captures most failures identified by humans (67 of 87) and surfaces 51 additional issues. As shown in Figure~\ref{fig:failure_breakdown}, most of the additional failures are concentrated in reproducibility and generalizability, which are more time-consuming for humans to evaluate. This pattern suggests execution enables \mechevalagent to uncover more issues, especially execution ones.


\para{Ablation highlights the problem of narrative-alone evaluation (Figure~\ref{fig:ablation}).}
Both ablated evaluators agree with humans far less than the full \mechevalagent pipeline, especially on execution dependent dimensions. This indicates that an AI evaluator without execution access and limited to narrative review is less reliable. In these settings, the evaluator also tends to over-assign FAIL as shown in Figure~\ref{fig:failure_all_metrics}, suggesting that execution provides needed grounding for calibrated judgments.

\subsection{A Closer Look at Failures and Disagreements}\label{subsec:closer_failure}

We now closely look into common failures identified by \mechevalagent and human experts, and analyze disagreements between them.

\begin{figure}[t]
    \centering
    \includegraphics[width=0.47\textwidth]{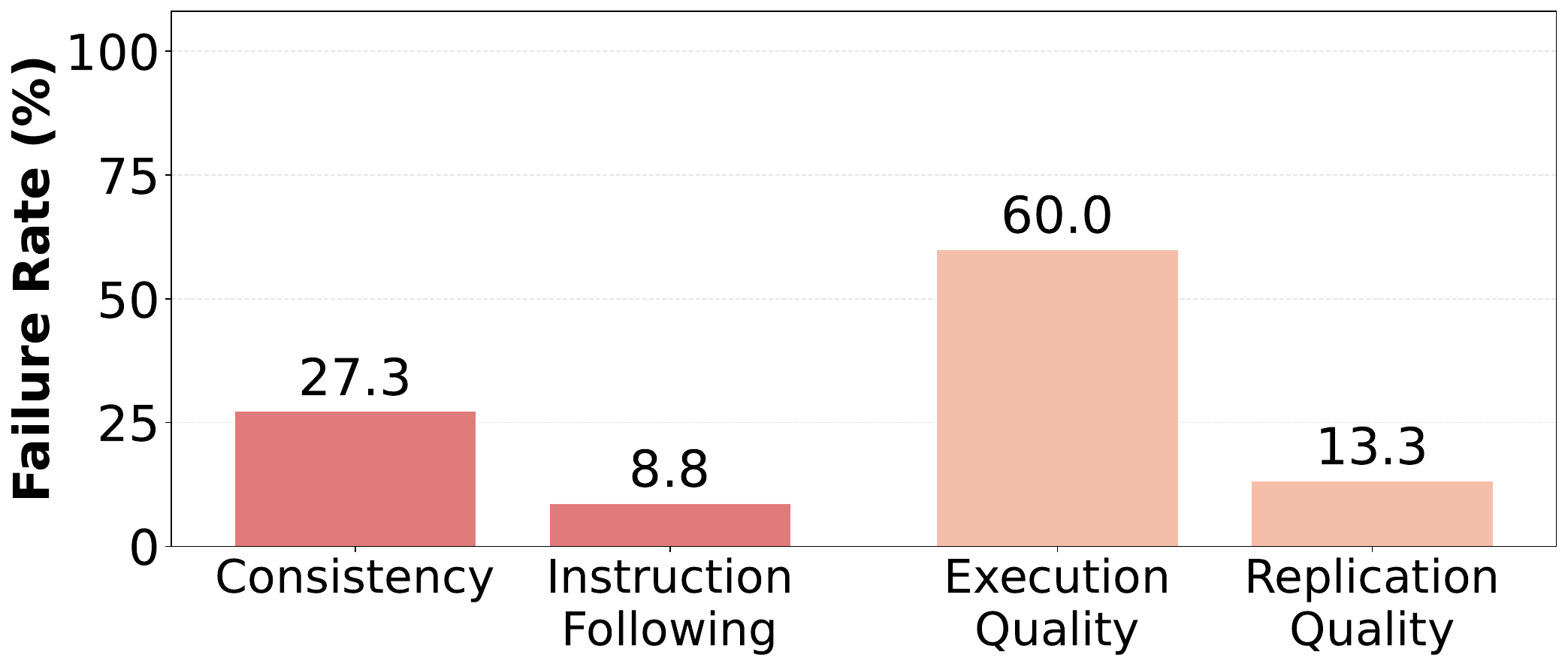}
    \includegraphics[width=0.3\textwidth]{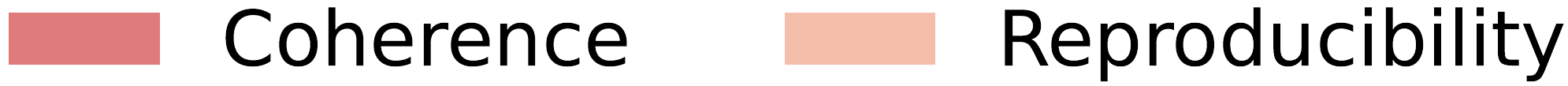}
    \caption{Average failure rates by subgroup. Consistency checks fail 27.3\% on average, and execution quality checks fail 60\% of the time.}
    \label{fig:failure_subgroup}
\end{figure}
\para{Execution failures and weak reproducibility remain common.}
Zooming in the high reproducibility failures observed in Figure~\ref{fig:and_result},
Figure~\ref{fig:failure_subgroup} shows that execution failures are more common. 
The high failure rate in execution evaluation reflects difficulties with missing packages and transformer internals, often showing up as shape errors in residual streams and attention heads (e.g., ``head-level activation patching mismatch'' from a replication task, and ``API compatibility issue'' from a human repository). 
In agent-generated research, these errors often appear early and are fixed in subsequent blocks. For example, after encountering a shape mismatch or missing key, the agent may inspect tensor shapes or enumerate available keys to find the correct one. 

However, a more serious problem is that results are sometimes not reproducible, undermining confidence in the conclusions.
In \texttt{acronyms}, a replication task that seeks a mechanistic account of predicting multiple consecutive tokens (Appendix~\ref{app:sub_experiment_setup}), 
the research agent's report claims to use the logit correlation metric to evaluate the circuit. However, \mechevalagent found that during replication, though the identified circuits overlap, logit correlation metrics deviate by more than 8\% from the original (Original value: 0.66, Replication value: 0.72), and this discrepancy appears across all three evaluation runs. Although the replicated results have better performance than the original, it reveals concerns in validity of the original method and the evaluation procedure. Inspecting the cause shown in Table~\ref{tab:issue_examples}, we found that the research agent evaluated correlation on only a subset of examples without explicitly noting this choice (first 20 examples), whereas \mechevalagent followed the stated instructions and evaluated on the full dataset. 
This observation point to issues in the original evaluation procedure and raise concerns about reproducibility. 

We observe a related issue in \texttt{erasing}~\cite{gandikota2024elm}, a human-written repository that proposes an approach to concept-level unlearning, with more detailed description in Appendix~\ref{app:sub_experiment_setup}. Here, \mechevalagent found that probe-training code and adversarial testing scripts are missing. Although the work primarily focuses on effective unlearning, the sections on probe and activation analysis and robustness to attacks serve as important evaluation components. While this failure is captured by our checklist item on consistency between stated experiments and implemented code, it also directly limits reproducibility because a user cannot reconstruct the full set of evaluation. Beyond missing evaluation scripts, \mechevalagent found that \texttt{belief}~\cite{prakash2025languagemodelsuselookbacks}, another human-written repository that studies how LMs represent characters' beliefs (Appendix~\ref{app:sub_experiment_setup}), contains an invalid Jupyter notebook file related to their BigToM causal model experiments.

\para{Narratives of agent-generated research fail to be consistent and lack sufficient justification.}
With respect to \emph{coherence} failures, Figure~\ref{fig:failure_subgroup} shows that the average failure rate in consistency subgroups can reach 27.3\%. As shown in Figure~\ref{fig:failure_all_metrics}, the most common consistency failure is missing statistical significance. 
We also observe failure rates of 23.3\% for plan-implementation consistency (\texttt{CS2}) and 6.7\% for effect size (\texttt{CS3}).

A particularly concerning failure is the lack of sufficient justification which has 23.3\% failure rate (Figure~\ref{fig:failure_all_metrics}). For example, as shown in Table~\ref{tab:issue_examples}, on the \texttt{ioi} task, which is a replication task focus on mechanistically understand the indirect object identification task (Appendix~\ref{app:sub_experiment_setup}), \mechevalagent noted that a ``strongly supports'' conclusion was drawn despite a -4.2\% circuit performance, which contradicts the reported evidence. These failures suggest potential hallucination in the research agent, possibly due to memorization of existing work in training data.

\begin{table*}[t]
\centering
\small
\caption{Examples of issues identified by \mechevalagent (\raisebox{-0.1em}{\includegraphics[height=1em]{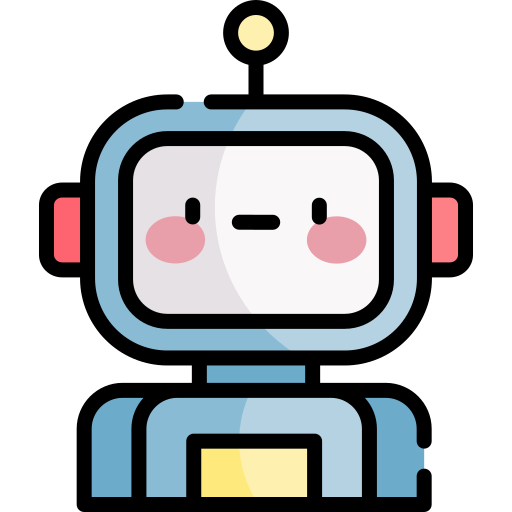}}) and human experts (\raisebox{-0.1em}{\includegraphics[height=1em]{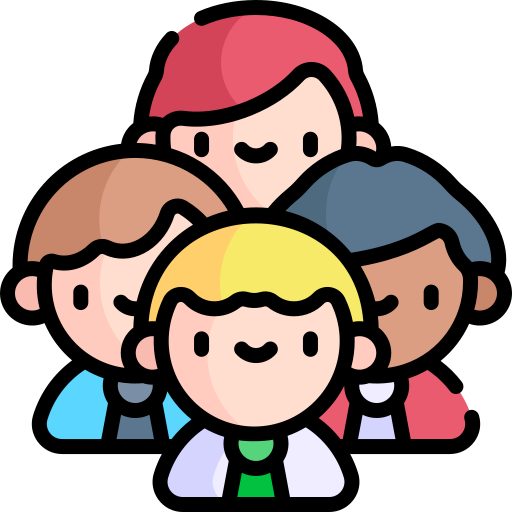}}). \mechevalagent provides specific, execution-grounded rationales, while human rationales tend to be more general. \textbf{R} = research agent generated repository, \textbf{H} = human-written repository. The detailed description of the tasks are in Appendix~\ref{app:sub_experiment_setup}.}
\label{tab:issue_examples}
\begin{tabular}{p{8cm}p{8cm}}
\toprule
\textbf{Code or Claims in Repository} & \textbf{Issues Identified} \\
\midrule
\textbf{[R]}
\texttt{for ex in examples\colorbox{yellow}{[:20]}: \#code for calculating the logic correlation metric}
[Source: \texttt{acronyms}: to understand multiple consecutive token predictions]
& \raisebox{-0.1em}{\includegraphics[height=1em]{plots/others/robot_6134346.png}} Logit correlation metrics deviate by more than 8\% during evaluation. (\texttt{DE1})
\newline
\newline
\raisebox{-0.1em}{\includegraphics[height=1em]{plots/others/candidates_12983796.png}} N/A \\
\midrule
\textbf{[R]} \texttt{logic\_diff\_retain
=\colorbox{yellow}{circuit\_diffs.mean()}/
\colorbox{yellow}{baseline\_diff}*100} logic\_diff = -4.2\% 
\newline
[Source: \texttt{ioi}: to understand indirect object identification]
& \raisebox{-0.1em}{\includegraphics[height=1em]{plots/others/robot_6134346.png}} The circuit verification shows -4.2\% performance retention. (\texttt{CS4}) 
\newline
\raisebox{-0.1em}{\includegraphics[height=1em]{plots/others/candidates_12983796.png}} The justification of getting the result is insufficient \\
\midrule
\textbf{[H]} No related code presented in the repository. In their paper, they reported probe analysis (Section:Probing and Activation Analysis) and tested the robustness of the attack (Section: Robustness to Attacks).
[Source: \href{https://github.com/rohitgandikota/erasing-llm}{\texttt{erasing}}~\cite{gandikota2024elm} Accessed: Jan 27, 2026]
& \raisebox{-0.1em}{\includegraphics[height=1em]{plots/others/robot_6134346.png}} 2 experiments are missing: (1) no attack code exists despite results being reported. (2) no probing or activation analysis code. (\texttt{CS2}) \newline
\raisebox{-0.1em}{\includegraphics[height=1em]{plots/others/candidates_12983796.png}} There are parts of the results that we don't see implemented \\
\midrule
\textbf{[H]} This \href{https://github.com/Nix07/belief_tracking/blob/main/notebooks/bigToM/causalmodel_exps.ipynb}{notebook} file has syntax error and is an invalid notebook.
[Source: \href{https://github.com/Nix07/belief_tracking}{\texttt{belief}}~\cite{prakash2025languagemodelsuselookbacks} Accessed: Jan 27, 2026]
& \raisebox{-0.1em}{\includegraphics[height=1em]{plots/others/robot_6134346.png}} \texttt{causalmodel\_exps.ipynb} file is not runnable (\texttt{C1}) \newline
\raisebox{-0.1em}{\includegraphics[height=1em]{plots/others/candidates_12983796.png}} \texttt{causalmode\_exps.ipynb} could not be opened because it doesn't finish closing the curly braces of the json file. \\
\bottomrule
\end{tabular}
\label{tab:venn_example}
\end{table*}

\para{\mechevalagent surfaces issues humans overlook.} 
Table~\ref{tab:venn_example} shows that \mechevalagent tends to provide more detailed and specific rationales than human experts, making the feedback more actionable for improving research agents. Human rationales are typically more general, likely due to limited time for thorough inspection. In \texttt{unanswerable}, an open-ended question task that explains how models solve unanswerable questions (Appendix~\ref{app:sub_experiment_setup}), \mechevalagent flagged that the choice of final circuit size lacks explicit justification and that validation results showing circuit ablation being only 0.73$\times$ as impactful as random ablation directly contradict the claimed importance. In contrast, human judges only noted that the justification was insufficient.

Figure~\ref{fig:failure_breakdown}  further shows that \mechevalagent surfaces more unique issues across all dimensions: 22.6\% in coherence, 55.3\% in reproducibility, and 57.1\% in generalizability. While some differences arise from interpretation, the agent consistently identifies execution-related problems. For example, as discussed earlier, in \texttt{acronyms}, \mechevalagent found that the logit correlation metric deviates substantially,
whereas human judges did not identify any issues.

Human evaluation can also be affected by limited resources or domain knowledge, and \mechevalagent helps compensate for these constraints. In \texttt{uncertainty}, an open-ended question task that aims to identify circuits related to uncertainty (Appendix~\ref{app:sub_experiment_setup}), human judges criticized the agent under \texttt{CS4} for switching to GPT-2 Medium without justification. However, code inspection reveals an \texttt{if} condition that switches models only after a load failure. By executing and inspecting the code, \mechevalagent provides a more accurate assessment.

Conversely, the 20 issues identified exclusively by human experts show that human evaluation still adds valuable insights beyond \mechevalagent. For example, in \texttt{multilingual}, an open-ended question task that aims to understand how models ``think'' under multilingual instructions (Appendix~\ref{app:sub_experiment_setup}), human judges identified a hallucination where the reported support value of $+$0.13 for an identified neuron was actually derived only from the single translation task, rather than from the average performance across the full task.

\para{Disagreements stem from different interpretations of checklists.}
Beyond surfacing issues, \mechevalagent shows high agreement with human evaluators, including both issues and pass cases. However, there are still some disagreements, especially in reproducibility and generalizability (Figure~\ref{fig:ablation}).
Figure~\ref{subfig:and_vs_all} shows that much of this comes from the Redundancy check (\texttt{C3}). Our checklist specifies that code adding new information should not be considered redundant. However, when the agent copies the same block repeatedly to run on different inputs instead of writing reusable functions, our evaluators mark it as redundant while human experts do not. This disagreement highlights inefficiencies in agent code generation, which is common among coding agents~\cite{horikawa2025agentic,azeem2025ai}. The high rated quality on this evaluation also supports that the evaluation is reasonable.

Generalizability metrics also show disagreement. For method generalizability (\texttt{GT3}), humans and evaluators differ on what counts as a new method. For model generalizability (\texttt{GT1}), disagreement arises from evaluation depth. Due to different model sizes, evaluation sometimes only assess high-level generalization. For example, the evaluation may simply check whether the function of early layers transfers to a new model, which human experts consider insufficient to demonstrate true generalization. Therefore, their rated quality falls mostly into Agree category.

\para{Efficiency comparison.} In terms of efficiency, to perform agreement evaluation, we asked human  judges to use the exact same checklists and instructions as the agents to assess the research, including reproducing the repository and testing generalizability. They spent an average of 2.2 hours per evaluation task, with evaluations of human-written repositories often exceeding 3.35 hours, as shown in Table~\ref{tab:efficiency}. 
In contrast, \mechevalagent typically completes evaluations in under 30 minutes for agent generated tasks and around one hour for human-written repositories.

Overall, \mechevalagent raises the floor of evaluation quality by efficiently surfacing issues that humans may overlook, while human expertise remains valuable for catching edge cases that automated evaluation misses.

\section{Related Work}

\para{Agents for Autonomous Research.}
LLM-based agents now automate research across multiple stages. Some focus on ideation and hypothesis generation~\cite{zhou2024hypothesis,baek2025researchagent,gottweis2025towards}, others on execution-grounded discovery through iterative code and environment feedback~\cite{novikov2025alphaevolve,jiang2025aide}. End-to-end systems close the loop from ideation to experimentation and writing, producing papers and repositories with code~\cite{lu2024ai,yamada2025ai,schmidgall2025agent,ifargan2025autonomous,jansen2025codescientist}. However, these systems produce heterogeneous outputs, complicating comparison. Our work standardizes research agent outputs into a unified trace that enables evaluation beyond narrative review alone.

\para{Evaluating Research.}
Recent efforts evaluate research agents along different dimensions. LLM-based peer review assesses paper narratives: \citet{liang2024can} found GPT-4 reviews overlap with human reviews at rates comparable to inter-reviewer agreement, and \citet{zhuang2025llmreview} survey LLM capabilities for checklist verification and error detection. However, these approaches treat code as readable but not executable. Benchmarks have begun addressing this gap: ResearchRubrics~\cite{sharma2025researchrubricsbenchmarkpromptsrubrics} and DeepResearch Bench~\cite{du2025deepresearchbenchcomprehensivebenchmark} assess long-form written outputs, while Exp-Bench~\cite{kon2025expbenchaiconductai} extends evaluation to code execution and experimental correctness. Other efforts benchmark ideation and hypothesis generation~\cite{guo2025ideabench,liu2025hypobench}. While these approaches advance grounded evaluation, they largely focus on reproducing a given experiment, with only access to the paper, which is unfair replication tasks. Our execution-grounded assessment framework is the first framework to evaluate this external validity, while also verifying that conclusions are grounded in actual execution rather than narrative alone. 
\section{Concluding Discussion} \label{sec:discuss}

Our work aims to verify the science beyond reviewing the paper.
To do that, we propose a novel evaluation framework consisting of coherence, reproducibility, and generalizability.
We further develops \mechevalagent and uses mechanistic interpretability as a testbed to compare agent-based evaluation with human evaluation. 
Our approach achieves above 80\% agreement with human experts while surfacing 51 additional issues that humans miss. 
Our results demonstrate that AI agents can raise the floor of evaluating scientific research by exposing concrete, execution-grounded issues that humans often overlook.

\para{From Narrative to Execution.}
We advocate a shift from narrative-alone review toward integrating execution-grounded evaluation into research assessment to improve rigor and reliability in scientific research.
Execution is crucial for research evaluation, yet reviewers often skip it due to time constraints and incomplete replication instructions. Our results point toward a natural division of labor. Human reviewers excel at judging novelty, contribution, and framing, areas where AI is 10 times less likely to comment~\cite{liang2024can}. The 20 issues identified exclusively by humans in our study involve interpretation and scope judgments requiring broader context. Automated evaluation, by contrast, excels at execution-heavy checks that humans skip under deadline pressure: verifying that code runs, that outputs match claims, and that results replicate. These are precisely the areas where \mechevalagent surfaces issues that humans miss.

Passing \mechevalagent does not guarantee high quality, but failing it highlights concrete, actionable weaknesses. Moreover, \mechevalagent pinpoints what needs revision and provide specific, execution-grounded rationales, while
human judges often give general rationales like ``insufficiently justified''.
This aligns with \citet{bahammam2025peer}, who argue that AI should support human editors by catching basic errors before external review. 

Beyond evaluation, \mechevalagent's feedback can help identify hallucinations in research agents and improve their reliability. As discussed in Section~\ref{subsec:closer_failure}, research agents often report positive results despite weak evidence and exploit ambiguity to make their outputs appear successful. Execution-grounded evaluation catches these issues early, before they propagate into published claims.

In short, we call for continued efforts to build evaluation frameworks that integrate narrative review with execution-grounded checks, moving research assessment toward verifying the science itself rather than just the story.

\para{Challenges in Agent-Based Evaluation.}
To proceed, we note three particular challenges. First, binary checklists can be brittle when evidence is mixed. They force hard decisions where graded judgments would better reflect uncertainty, making outcomes sensitive to borderline cases and underspecified scenarios. Because research is highly open-ended, this underspecified space may be larger than in other domains. 
Second, we observe occasional optimistic behavior in ambiguous settings. In early runs, the evaluation agent sometimes modifies source files to justify a PASS despite prompts forbidding edits, or downplays negative evidence under underspecified failure handling. Although introducing stricter prompts nearly mitigates these issues, they highlight the need for caution in agent-based evaluation. More broadly, evaluation agents may default to neutral or positive judgments when interpretation is unclear. Binary checklists expose this tendency by forcing categorical decisions, motivating clearer guidance for handling uncertainty in automated evaluation.
The third challenge arises from limited instruction-following ability. As noted above, even when we explicitly forbid modifications to source files, the agent sometimes still makes changes. When evaluating code quality, the model is instructed to copy and re-run the exact code, yet it occasionally executes specific functions instead or silently alters the copied code. A related issue arises in metric definitions. For example, although the prompt clearly states that fixing a previous error should not count as redundancy, the model sometimes marks such cases as redundant, even when its own rationale acknowledges the change fixes an earlier error. These behaviors highlight limitations in instruction following that remain a challenge for coding agents. Despite these challenges, \mechevalagent achieves above 80\% agreement with human evaluators, demonstrating the viability of automated research evaluation. We encourage continued efforts to improve instruction following and uncertainty handling in evaluation agents.

\para{Limitations.} Our current implementation uses Claude Code as the sole evaluation agent. While our results demonstrate strong alignment with human judgment, future work could explore ensembles of different models to further improve robustness. Additionally, due to resource constraints, each repository was reviewed by a single human expert rather than multiple independent reviewers, and human evaluation was conducted on the first evaluation run rather than all three runs. Despite these constraints, the high agreement between our pipeline and human experts suggests that our findings are reliable, and we expect that additional human reviewers would further validate the patterns we observe.

\section*{Acknowledgements}
We gratefully thank Yonatan Belinkov, Tal Haklay, Austin Kozlowski, and Tamar Rott Shaham for fruitful discussions. We also thank Modal for generously providing computing credits.
\section*{Impact Statement}
Our work proposes a new evaluation framework that combines narrative review with execution-grounded checks to make research evaluation more rigorous. More broadly, our work aims to advance research evaluation toward more scalable and automated processes. This effort is not limited to machine learning and can be generalized to other fields. This research does not present specific ethical concerns or societal implications beyond those.

\bibliography{main}

@misc{anthropic2025claudecode,
  author       = {{Anthropic}},
  title        = {{Claude Code: an agentic coding assistant}},
  howpublished = {\url{https://code.claude.com/docs/en/sub-agents}},
  year         = {2025},
  note         = {Agentic coding workflows and subagent framework for automated code generation and execution},
}

@misc{openai2025codex,
  author       = {{OpenAI}},
  title        = {{OpenAI Codex: cloud-based software engineering agent}},
  howpublished = {\url{https://openai.com/index/introducing-codex/}},
  year         = {2025},
  note         = {Software engineering agent that can write, test, and deploy code autonomously},
}

@misc{sharma2025researchrubricsbenchmarkpromptsrubrics,
      title={ResearchRubrics: A Benchmark of Prompts and Rubrics For Evaluating Deep Research Agents}, 
      author={Manasi Sharma and Chen Bo Calvin Zhang and Chaithanya Bandi and Clinton Wang and Ankit Aich and Huy Nghiem and Tahseen Rabbani and Ye Htet and Brian Jang and Sumana Basu and Aishwarya Balwani and Denis Peskoff and Marcos Ayestaran and Sean M. Hendryx and Brad Kenstler and Bing Liu},
      year={2025},
      eprint={2511.07685},
      archivePrefix={arXiv},
      primaryClass={cs.AI},
      url={https://arxiv.org/abs/2511.07685}, 
}

@misc{kon2025expbenchaiconductai,
      title={EXP-Bench: Can AI Conduct AI Research Experiments?}, 
      author={Patrick Tser Jern Kon and Jiachen Liu and Xinyi Zhu and Qiuyi Ding and Jingjia Peng and Jiarong Xing and Yibo Huang and Yiming Qiu and Jayanth Srinivasa and Myungjin Lee and Mosharaf Chowdhury and Matei Zaharia and Ang Chen},
      year={2025},
      eprint={2505.24785},
      archivePrefix={arXiv},
      primaryClass={cs.AI},
      url={https://arxiv.org/abs/2505.24785}, 
}

@misc{du2025deepresearchbenchcomprehensivebenchmark,
      title={DeepResearch Bench: A Comprehensive Benchmark for Deep Research Agents}, 
      author={Mingxuan Du and Benfeng Xu and Chiwei Zhu and Xiaorui Wang and Zhendong Mao},
      year={2025},
      eprint={2506.11763},
      archivePrefix={arXiv},
      primaryClass={cs.CL},
      url={https://arxiv.org/abs/2506.11763}, 
}

@article{camerer2018replicability,
  title   = {Evaluating the replicability of social science experiments in Nature and Science between 2010 and 2015},
  author  = {Camerer, Colin F. and Dreber, Anna and Holzmeister, Felix and Ho, Teck-Hua and Huber, J{\"u}rgen and Johannesson, Magnus and Kirchler, Michael and others},
  journal = {Nature Human Behaviour},
  volume  = {2},
  number  = {9},
  pages   = {637--644},
  year    = {2018}
}

@misc{lazic2025internalreplicationtoolevaluating,
      title={Internal replication as a tool for evaluating reproducibility in preclinical experiments}, 
      author={Stanley E. Lazic},
      year={2025},
      eprint={2506.03468},
      archivePrefix={arXiv},
      primaryClass={stat.AP},
      url={https://arxiv.org/abs/2506.03468}, 
}

@article{errington2021replicability,
  title   = {Investigating the replicability of preclinical cancer biology},
  author  = {Errington, Timothy M. and Mathur, Maya and Soderberg, Courtney K. and Denis, Alexandria and Perfito, Nicole and Iorns, Elizabeth and Nosek, Brian A.},
  journal = {eLife},
  volume  = {10},
  pages   = {e71601},
  year    = {2021}
}

@software{goodfire2025scribe,
  author       = {{Goodfire AI}},
  title        = {{Scribe: Jupyter Server + Notebooks for CLI Agents}},
  year         = {2025},
  url          = {https://github.com/goodfire-ai/scribe},
  note         = {GitHub repository; gives CLI agents access to Jupyter servers and automatically records code and outputs in notebooks}  
}

@misc{kim2025llmspursueagenticinterpretability,
      title={Because we have LLMs, we Can and Should Pursue Agentic Interpretability}, 
      author={Been Kim and John Hewitt and Neel Nanda and Noah Fiedel and Oyvind Tafjord},
      year={2025},
      eprint={2506.12152},
      archivePrefix={arXiv},
      primaryClass={cs.AI},
      url={https://arxiv.org/abs/2506.12152}, 
}

@inproceedings{feucht2025arithmetic,
  title={Vector Arithmetic in Concept and Token Subspaces},
  author={Sheridan Feucht and Byron Wallace and David Bau},
  booktitle={Second Mechanistic Interpretability Workshop at NeurIPS},
  year={2025},
  url={https://arithmetic.baulab.info}
}

@inproceedings{todd2024function,
    title={Function Vectors in Large Language Models}, 
    author={Eric Todd and Millicent L. Li and Arnab Sen Sharma and Aaron Mueller and Byron C. Wallace and David Bau},
    booktitle={Proceedings of the 2024 International Conference on Learning Representations},
    url={https://openreview.net/forum?id=AwyxtyMwaG},
    note={arXiv:2310.15213},
    year={2024},
}

@misc{prakash2025languagemodelsuselookbacks,
      title={Language Models use Lookbacks to Track Beliefs}, 
      author={Nikhil Prakash and Natalie Shapira and Arnab Sen Sharma and Christoph Riedl and Yonatan Belinkov and Tamar Rott Shaham and David Bau and Atticus Geiger},
      year={2025},
      eprint={2505.14685},
      archivePrefix={arXiv},
      primaryClass={cs.CL},
      url={https://arxiv.org/abs/2505.14685}, 
}

@article{sensharma2023filter,
    title={LLMs Process Lists With General Filter Heads}, 
    author={Arnab Sen Sharma and Giordano Rogers and Natalie Shapira and David Bau},
    year={2025},
    eprint={2510.26784},
    archivePrefix={arXiv},
    primaryClass={cs.CL}
}

@article{tan2025interpdetect,
  title={InterpDetect: Interpretable Signals for Detecting Hallucinations in Retrieval-Augmented Generation},
  author={Tan, Likun and Huang, Kuan-Wei and Shi, Joy and Wu, Kevin},
  journal={arXiv preprint arXiv:2510.21538},
  year={2025}
}

@misc{sandmann2025iterativeinferencechessplayingneural,
      title={Iterative Inference in a Chess-Playing Neural Network}, 
      author={Elias Sandmann and Sebastian Lapuschkin and Wojciech Samek},
      year={2025},
      eprint={2508.21380},
      archivePrefix={arXiv},
      primaryClass={cs.LG},
      url={https://arxiv.org/abs/2508.21380}, 
}

@article{gandikota2024elm,
  title={Erasing Conceptual Knowledge from Language Models},
  author={Rohit Gandikota and Sheridan Feucht and Samuel Marks and David Bau},
  journal={arXiv preprint arXiv:2410.02760},
  year={2024}
}

@inproceedings{hernandez2024linearity,
    title={Linearity of Relation Decoding in Transformer Language Models}, 
    author={Evan Hernandez and Arnab Sen Sharma and Tal Haklay and Kevin Meng and Martin Wattenberg and Jacob Andreas and Yonatan Belinkov and David Bau},
    booktitle={Proceedings of the 2024 International Conference on Learning Representations},
    year={2024},
}

@article{liu2025hypobench,
  title={Hypobench: Towards systematic and principled benchmarking for hypothesis generation},
  author={Liu, Haokun and Huang, Sicong and Hu, Jingyu and Zhou, Yangqiaoyu and Tan, Chenhao},
  journal={arXiv preprint arXiv:2504.11524},
  year={2025}
}

@article{lu2024ai,
  title={The ai scientist: Towards fully automated open-ended scientific discovery},
  author={Lu, Chris and Lu, Cong and Lange, Robert Tjarko and Foerster, Jakob and Clune, Jeff and Ha, David},
  journal={arXiv preprint arXiv:2408.06292},
  year={2024}
}

@article{gurnee2024universal,
  title={Universal neurons in gpt2 language models},
  author={Gurnee, Wes and Horsley, Theo and Guo, Zifan Carl and Kheirkhah, Tara Rezaei and Sun, Qinyi and Hathaway, Will and Nanda, Neel and Bertsimas, Dimitris},
  journal={arXiv preprint arXiv:2401.12181},
  year={2024}
}

@article{meng2022locating,
  title={Locating and Editing Factual Associations in {GPT}},
  author={Kevin Meng and David Bau and Alex Andonian and Yonatan Belinkov},
  journal={Advances in Neural Information Processing Systems},
  volume={35},
  year={2022}
}

@misc{kozlov_persona_collapse,
  title        = {Persona Collapse},
  author       = {Alexey Kozlov},
  year         = {2025},
  howpublished = {\url{https://github.com/akozlo/Persona-Collapse-blog}},
  note         = {Accessed: 2026-01-21}
}

@article{horikawa2025agentic,
  title={Agentic Refactoring: An Empirical Study of AI Coding Agents},
  author={Horikawa, Kosei and Li, Hao and Kashiwa, Yutaro and Adams, Bram and Iida, Hajimu and Hassan, Ahmed E},
  journal={arXiv preprint arXiv:2511.04824},
  year={2025}
}

@article{azeem2025ai,
  title={AI vs. Human Programmers: Complexity and Performance in Code Generation},
  author={Azeem, Samina and Naveed, Muhammad Shumail and Sajid, Muhammad and Ali, Imran},
  journal={VAWKUM Transactions on Computer Sciences},
  volume={13},
  number={1},
  pages={201--216},
  year={2025}
}

@article{zhou2024hypothesis,
  title={Hypothesis generation with large language models},
  author={Zhou, Yangqiaoyu and Liu, Haokun and Srivastava, Tejes and Mei, Hongyuan and Tan, Chenhao},
  journal={arXiv preprint arXiv:2404.04326},
  year={2024}
}

@inproceedings{baek2025researchagent,
  title={Researchagent: Iterative research idea generation over scientific literature with large language models},
  author={Baek, Jinheon and Jauhar, Sujay Kumar and Cucerzan, Silviu and Hwang, Sung Ju},
  booktitle={Proceedings of the 2025 Conference of the Nations of the Americas Chapter of the Association for Computational Linguistics: Human Language Technologies (Volume 1: Long Papers)},
  pages={6709--6738},
  year={2025}
}

@inproceedings{guo2025ideabench,
  title={Ideabench: Benchmarking large language models for research idea generation},
  author={Guo, Sikun and Shariatmadari, Amir Hassan and Xiong, Guangzhi and Huang, Albert and Kim, Myles and Williams, Corey M and Bekiranov, Stefan and Zhang, Aidong},
  booktitle={Proceedings of the 31st ACM SIGKDD Conference on Knowledge Discovery and Data Mining V. 2},
  pages={5888--5899},
  year={2025}
}

@article{novikov2025alphaevolve,
  title={AlphaEvolve: A coding agent for scientific and algorithmic discovery},
  author={Novikov, Alexander and V{\~u}, Ng{\^a}n and Eisenberger, Marvin and Dupont, Emilien and Huang, Po-Sen and Wagner, Adam Zsolt and Shirobokov, Sergey and Kozlovskii, Borislav and Ruiz, Francisco JR and Mehrabian, Abbas and others},
  journal={arXiv preprint arXiv:2506.13131},
  year={2025}
}

@misc{bahammam2025peer,
  title={Peer review in the artificial intelligence era: A call for developing responsible integration guidelines},
  author={BaHammam, Ahmed Salem},
  journal={Nature and Science of Sleep},
  pages={159--164},
  year={2025},
  publisher={Taylor \& Francis}
}

@article{desai2025reproducibility,
  title={What is reproducibility in artificial intelligence and machine learning research?},
  author={Desai, Abhyuday and Abdelhamid, Mohamed and Padalkar, Nakul R},
  journal={AI Magazine},
  volume={46},
  number={2},
  pages={e70004},
  year={2025},
  publisher={Wiley Online Library}
}

@article{haibe2020transparency,
  title={Transparency and reproducibility in artificial intelligence},
  author={Haibe-Kains, Benjamin and Adam, George Alexandru and Hosny, Ahmed and Khodakarami, Farnoosh and Massive Analysis Quality Control (MAQC) Society Board of Directors Shraddha Thakkar 35 Kusko Rebecca 36 Sansone Susanna-Assunta 37 Tong Weida 35 Wolfinger Russ D. 38 Mason Christopher E. 39 Jones Wendell 40 Dopazo Joaquin 41 Furlanello Cesare 42 and Waldron, Levi and Wang, Bo and McIntosh, Chris and Goldenberg, Anna and Kundaje, Anshul and others},
  journal={Nature},
  volume={586},
  number={7829},
  pages={E14--E16},
  year={2020},
  publisher={Nature Publishing Group UK London}
}

@article{beam2020challenges,
  title={Challenges to the reproducibility of machine learning models in health care},
  author={Beam, Andrew L and Manrai, Arjun K and Ghassemi, Marzyeh},
  journal={Jama},
  volume={323},
  number={4},
  pages={305--306},
  year={2020},
  publisher={American Medical Association}
}

@misc{GPTZero_NeurIPS_Hallucination_Check_2026,
  title        = {GPTZero finds 100 new hallucinations in NeurIPS 2025 accepted papers},
  author       = {Nazar Shmatko and Alex Adam and Paul Esau},
  year         = {2026},
  howpublished = {Online},
  month        = {January 21},
  note         = {Accessed: 2026-01-23},
  url          = {https://tinyurl.com/4n2h9jfk}
}

@article{lee2025role,
  title={The role of large language models in the peer-review process: opportunities and challenges for medical journal reviewers and editors},
  author={Lee, Jisoo and Lee, Jieun and Yoo, Jeong-Ju},
  journal={Journal of Educational Evaluation for Health Professions},
  volume={22},
  year={2025},
  publisher={Korea Health Personnel Licensing Examination Institute}
}

@article{liang2024can,
  title={Can large language models provide useful feedback on research papers? A large-scale empirical analysis},
  author={Liang, Weixin and Zhang, Yuhui and Cao, Hancheng and Wang, Binglu and Ding, Daisy Yi and Yang, Xinyu and Vodrahalli, Kailas and He, Siyu and Smith, Daniel Scott and Yin, Yian and others},
  journal={NEJM AI},
  volume={1},
  number={8},
  pages={AIoa2400196},
  year={2024},
  publisher={Massachusetts Medical Society}
}

@misc{PaperReviewAI_Stanford_2025,
  title        = {PaperReview.ai: Stanford Agentic Reviewer for AI-Assisted Paper Feedback},
  author       = {{Stanford ML Group}},
  year         = {2025},
  howpublished = {Online},
  note         = {Accessed: 2026-01-23},
  url          = {https://paperreview.ai/}
}

@article{yu2024researchtown,
  title={Researchtown: Simulator of human research community},
  author={Yu, Haofei and Hong, Zhaochen and Cheng, Zirui and Zhu, Kunlun and Xuan, Keyang and Yao, Jinwei and Feng, Tao and You, Jiaxuan},
  journal={arXiv preprint arXiv:2412.17767},
  year={2024}
}

@online{RetractionWatch2024Psychology,
  author       = {Lori Youmshajekian},
  title        = {Exclusive: Psychology researcher loses PhD after allegedly using husband in study and making up data},
  year         = {2024},
  month        = apr # "~26",
  url          = {https://tinyurl.com/32tcj86c},
  note         = {Accessed: 2026-01-25},
  organization = {Retraction Watch}
}

@online{DataColada2024GinoReport,
  author       = {Joe Simmons},
  title        = {Harvard’s Gino Report Reveals How A Dataset Was Altered},
  year         = {2024},
  month        = jul # "~9",
  url          = {https://datacolada.org/118},
  note         = {Accessed: 2026-01-25},
  organization = {Data Colada}
}

@article{abogunrin2025much,
  title={How much can we save by applying artificial intelligence in evidence synthesis? Results from a pragmatic review to quantify workload efficiencies and cost savings},
  author={Abogunrin, Seye and Muir, Jeffrey M and Zerbini, Clarissa and Sarri, Grammati},
  journal={Frontiers in Pharmacology},
  volume={16},
  pages={1454245},
  year={2025},
  publisher={Frontiers Media SA}
}

@article{gottweis2025towards,
  title={Towards an AI co-scientist},
  author={Gottweis, Juraj and Weng, Wei-Hung and Daryin, Alexander and Tu, Tao and Palepu, Anil and Sirkovic, Petar and Myaskovsky, Artiom and Weissenberger, Felix and Rong, Keran and Tanno, Ryutaro and others},
  journal={arXiv preprint arXiv:2502.18864},
  year={2025}
}

@article{jiang2025aide,
  title={Aide: Ai-driven exploration in the space of code},
  author={Jiang, Zhengyao and Schmidt, Dominik and Srikanth, Dhruv and Xu, Dixing and Kaplan, Ian and Jacenko, Deniss and Wu, Yuxiang},
  journal={arXiv preprint arXiv:2502.13138},
  year={2025}
}

@article{yamada2025ai,
  title={The ai scientist-v2: Workshop-level automated scientific discovery via agentic tree search},
  author={Yamada, Yutaro and Lange, Robert Tjarko and Lu, Cong and Hu, Shengran and Lu, Chris and Foerster, Jakob and Clune, Jeff and Ha, David},
  journal={arXiv preprint arXiv:2504.08066},
  year={2025}
}

@article{schmidgall2025agent,
  title={Agent laboratory: Using llm agents as research assistants},
  author={Schmidgall, Samuel and Su, Yusheng and Wang, Ze and Sun, Ximeng and Wu, Jialian and Yu, Xiaodong and Liu, Jiang and Liu, Zicheng and Barsoum, Emad},
  journal={arXiv preprint arXiv:2501.04227},
  year={2025}
}

@article{ifargan2025autonomous,
  title={Autonomous llm-driven research—from data to human-verifiable research papers},
  author={Ifargan, Tal and Hafner, Lukas and Kern, Maor and Alcalay, Ori and Kishony, Roy},
  journal={NEJM AI},
  volume={2},
  number={1},
  pages={AIoa2400555},
  year={2025},
  publisher={Massachusetts Medical Society}
}

@inproceedings{jansen2025codescientist,
  title={Codescientist: End-to-end semi-automated scientific discovery with code-based experimentation},
  author={Jansen, Peter and Tafjord, Oyvind and Radensky, Marissa and Siangliulue, Pao and Hope, Tom and Dalvi, Bhavana and Majumder, Bodhisattwa Prasad and Weld, Daniel S and Clark, Peter},
  booktitle={Findings of the Association for Computational Linguistics: ACL 2025},
  pages={13370--13467},
  year={2025}
}

@article{zhuang2025llmreview,
  title={Large language models for automated scholarly paper review: A survey},
  author={Zhuang, Zhenzhen and Chen, Jiandong and Xu, Hongfeng and Jiang, Yuwen and Lin, Jialiang},
  journal={Information Fusion},
  volume={124},
  pages={103332},
  year={2025}
}

@inproceedings{wang2022interpretability,
  title={Interpretability in the Wild: a Circuit for Indirect Object Identification in {GPT}-2 Small},
  author={Wang, Kevin and Variengien, Alexandre and Conmy, Arthur and Shlegeris, Buck and Steinhardt, Jacob},
  booktitle={International Conference on Learning Representations},
  year={2023}
}

@article{hanna2023greaterthan,
  title={How does {GPT}-2 compute greater-than?: Interpreting mathematical abilities in a pre-trained language model},
  author={Hanna, Michael and Liu, Ollie and Variengien, Alexandre},
  journal={arXiv preprint arXiv:2305.00586},
  year={2023}
}

@article{olsson2022induction,
  title={In-context Learning and Induction Heads},
  author={Olsson, Catherine and Elhage, Nelson and Nanda, Neel and Joseph, Nicholas and DasSarma, Nova and Henighan, Tom and Mann, Ben and Askell, Amanda and Bai, Yuntao and Chen, Anna and others},
  journal={Transformer Circuits Thread},
  year={2022}
}

@inproceedings{nanda2023progress,
  title={Progress Measures for Grokking via Mechanistic Interpretability},
  author={Nanda, Neel and Chan, Lawrence and Lieberum, Tom and Smith, Jess and Steinhardt, Jacob},
  booktitle={International Conference on Learning Representations},
  year={2023}
}

@inproceedings{garcia2024does,
  title={How does gpt-2 predict acronyms? extracting and understanding a circuit via mechanistic interpretability},
  author={Garc{\'\i}a-Carrasco, Jorge and Mat{\'e}, Alejandro and Trujillo, Juan Carlos},
  booktitle={International Conference on Artificial Intelligence and Statistics},
  pages={3322--3330},
  year={2024},
  organization={PMLR}
}

@article{mcdougall2023copy,
  title={Copy Suppression: Comprehensively Understanding an Attention Head},
  author={McDougall, Callum and Conmy, Arthur and Rushing, Cody and McGrath, Thomas and Nanda, Neel},
  journal={arXiv preprint arXiv:2310.04625},
  year={2023}
}

@misc{heimersheim2023docstring,
  title={A Circuit for Python Docstrings in a 4-Layer Attention-Only Transformer},
  author={Heimersheim, Stefan and Janiak, Jett},
  year={2023},
  howpublished={Alignment Forum},
  url={https://tinyurl.com/ycynk5kv}
}

@misc{mathwin2023pronoun,
  title={Identifying a Preliminary Circuit for Predicting Gendered Pronouns in GPT-2 Small},
  author={Mathwin, Chris and Corlouer, Guillaume and Kran, Esben and Barez, Fazl and Nanda, Neel},
  year={2023},
  howpublished={MATS/Apart Mechanistic Interpretability Hackathon},
  url={https://itch.io/jam/mechint/rate/1889871}
}

@inproceedings{gross2024compact,
  title={Compact Proofs of Model Performance via Mechanistic Interpretability},
  author={Gross, Jason and Agrawal, Rajashree and Kwa, Thomas and Ong, Euan and Yip, Chun Hei and Gibson, Alex and Noubir, Soufiane and Chan, Lawrence},
  booktitle={Advances in Neural Information Processing Systems},
  volume={37},
  year={2024}
}

@misc{mcdougall2023brackets,
  title={Balanced Bracket Classifier: Mechanistic Interpretability Tutorial},
  author={McDougall, Callum},
  year={2023},
  howpublished={ARENA 3.0 Curriculum},
  url={https://arena3-chapter1-transformer-interp.streamlit.app/[1.5.1]_Balanced_Bracket_Classifier},
  note={Educational materials with documented circuit solution}
}

@article{kelly2014peer,
  title={Peer review in scientific publications: benefits, critiques, \& a survival guide},
  author={Kelly, Jacalyn and Sadeghieh, Tara and Adeli, Khosrow},
  journal={Ejifcc},
  volume={25},
  number={3},
  pages={227},
  year={2014}
}

@misc{papercopilot_neurips_statistics,
  title        = {NeurIPS Statistics},
  author       = {{PaperCopilot}},
  url          = {https://papercopilot.com/statistics/neurips-statistics/},
  note         = {Accessed: 2026-01-27}
}

@misc{neurips2025_review_reflections,
  title        = {Reflections on the 2025 Review Process from the Program Committee Chairs},
  author       = {{NeurIPS PC Chairs}},
  year         = {2025},
  url          = {https://tinyurl.com/mw48e9rs},
  note         = {Accessed: 2026-01-27}
}

@article{open2015estimating,
  title={Estimating the reproducibility of psychological science},
  author={Open Science Collaboration},
  journal={Science},
  volume={349},
  number={6251},
  pages={aac4716},
  year={2015},
  publisher={American Association for the Advancement of Science}
}

@article{nature_ai_peer_review_2025,
  title   = {AI is transforming peer review — and that’s raising concerns},
  author  = {Else, Holly},
  journal = {Nature},
  year    = {2025},
  doi     = {10.1038/d41586-025-02172-y},
  url     = {https://www.nature.com/articles/d41586-025-02172-y}
}
\bibliographystyle{icml2026}

\newpage
\appendix
\onecolumn

\section{Design and Experiment Setup Details}~\label{sec:app_checklist}In this section, we further discuss the details about the three dimensions we pick: coherence, reproducibility, and generalizability. Then, we provide the details of the items in the checklist and the tasks we pick.

\subsection{Design Details} \label{app:evaluator_details}This appendix provides additional details for the three evaluation dimensions used in \mechevalagent. The main paper presents a concise overview in Section~\ref{sec:method}.

\para{Coherence.}
Coherence evaluation includes \emph{consistency evaluation} and \emph{instruction following} evaluation.
\emph{Consistency evaluation} checks whether the agent's implementation, results, and conclusions match one another. It tests hallucinated findings by verifying that the implementation adheres to the stated plan, runs successfully, and produces the outputs referenced in the documentation. It also checks effect size, justification, and statistical significance. Together, these checks ensure that the reasoning chain from plan to code to analysis is logically sound and grounded in executable evidence.
Besides, we also evaluate \emph{instruction following}. This step assesses whether the agent performed the intended experiment. \mechevalagent has extra access to human input prompts to check whether it pursues the specified hypotheses, respects constraints, and follows the prescribed setup.

\para{Reproducibility.}
We evaluate two aspects of reproducibility: \emph{execution quality} and \emph{replication quality}. For execution quality, we check code runnability. However, runnable code is necessary but not sufficient for reproducibility, as it may still produce incorrect results due to over-claiming or hallucination.

For replication quality, we test whether the same model, in a fresh session and without access to the original report, can reproduce the reported results. This mirrors standard scientific practice, where reproducibility is established by independently re-running experiments~\cite{errington2021replicability,lazic2025internalreplicationtoolevaluating,camerer2018replicability}. We forbid access to the report to prevent implicit hallucination, where the model might copy conclusions regardless of actual results or reverse-engineer the replication from reported findings. The plan and implementation should contain all information needed to replicate. A separate replicator evaluator then verifies whether the same conclusions are reached. To further guard against hallucination, we check for references to external sources or reliance on memorized information not present in the repository. Successful replication provides stronger evidence than runnability alone, indicating that the experiment was well-specified, sufficiently documented, and free of hidden dependencies.

\para{Generalization.}
In this stage, the agent reads the research repository and evaluates whether the reported findings generalize beyond the original setting, such as to a new model, new data instances, or related tasks. The agent is allowed up to three trials to identify suitable alternative models or data. This constraint prevents unbounded search while still giving the agent enough flexibility to explore and produce stable, informative feedback.

\subsection{Experiment Setup Details.}\label{app:sub_experiment_setup}

\para{Checklist.} Across all settings, we use structured binary checklists to evaluate coherence, reproducibility, and generalizability, as shown in Table~\ref{tab:checklist}.
Each evaluator produces checklist-level judgments with concise rationales, along with a more detailed analysis report. The evaluation process, including execution and logs, is stored in a Jupyter notebook. Then it will generate a separate json file to summarize the results and the rationale. An example \mechevalagent output of consistency evaluation on \texttt{ioi} task is shown below:

\begin{verbatim}
  "Checklist": {
    "CS1_Results_vs_Conclusion": "FAIL",
    "CS2_Plan_vs_Implementation": "PASS",
    "CS3_Effect_Size": "FAIL",
    "CS4_Justification": "FAIL",
    "CS5_Statistical_Significance": "FAIL"
  },
  "Rationale": {
    "CS1_Results_vs_Conclusion": "The documentation claims the circuit
      'strongly supports' the hypothesis, but the actual circuit
      verification shows negative performance (-4.2% of baseline).",
    "CS2_Plan_vs_Implementation": "All five steps in the plan were
      implemented. While verification results were poor, the plan
      steps were executed as specified.",
    "CS3_Effect_Size": "The circuit verification shows -4.2% performance
      retention, meaning the circuit performs worse than random.",
    "CS4_Justification": "Multiple key choices lack justification:
      attention thresholds are arbitrary; the 'strongly supports'
      conclusion contradicts the -4.2% circuit performance.",
    "CS5_Statistical_Significance": "No error bars, confidence intervals,
      or statistical tests are reported."
  }
\end{verbatim}

\para{Example Input Prompt for Research Agents.}
Our research agent is built in Scribe and we design specific prompts to guide them through the tasks and regulate the output. In replication tasks, we give it more detailed research task, experiment setup, hypothesis to test, and possible method to use. While in open-ended tasks, we only give it the general questions and possible methodology. Here is an illustration of our input prompts. For more detailed prompts, please refer to supplementary material.

\begin{verbatim}
You are a **senior mechanistic interpretability researcher**.
### MODEL AND DATA ...
### GOAL
Identify a **precise circuit**—
a subset of attention heads and MLPs—that reproduces the model’s 
**sarcasm recognition behavior**.
The focus is on **how the model internally resolves conflict between 
literal content and intended meaning**.
...
### TASK DESCRIPTION
Sentences in the sarcasm dataset typically contain **conflicting cues** 
between surface meaning and pragmatic intent.
Example:...
Phenomena of interest may include (non-exhaustive):...
...
### SRC_NODES, CONSTRAINTS, EXPECTED OUTPUTS, FILES TO PRODUCE,
DOCUMENTATION REQUIREMENTS, OUTPUT SUMMARY 
...
\end{verbatim}

\para{Task Design.}
As we introduced in Section~\ref{sec:exp_setup}, we design three different categories of tasks including replication tasks, open-ended question tasks, and research done by human researchers. Replication tasks request the research agent to replicate an existing work, while open-ended questions are proposed by us and have not had specific research conducted on them. The research repositories of replication and open-ended tasks are all generated by an agent. In replication tasks, our input prompt to the research agent specifies the task, the hypothesis, and possible methodology (but not limited to those) they can use. In open-ended questions, our input prompt specifies the research question we are interested in and potential methods they can use to approach the question. But the agents are required to come up with their own hypothesis and method. The details of our task choices are shown in Table~\ref{tab:tasklist}. 
We provide additional details on each task category below.

\textit{Replication Tasks.} The replication tasks comprise ten well-established benchmarks from mechanistic interpretability literature, each with documented ground-truth circuits that enable systematic evaluation. 

\texttt{ioi} (Indirect Object Identification)~\citep{wang2022interpretability} requires predicting the indirect object in sentences with an ``ABB'' pattern (e.g., ``When John and Mary went to the store, John gave a drink to...'' $\rightarrow$ ``Mary''). This task isolates a 26-head circuit with seven functional classes, including Name Mover heads and S-Inhibition heads. \texttt{acronyms} (Acronym prediction)~\citep{garcia2024does} tests multi-token generation by predicting three-letter acronyms from expanded forms (e.g., ``Chief Executive Officer'' $\rightarrow$ ``CEO''). 
\texttt{greater\_than} (Greater-than)~\citep{hanna2023greaterthan} evaluates numerical reasoning with sentences like ``The war lasted from 1732 to 17...'' where the model must predict a valid year greater than the starting year. \texttt{pronoun} (Gendered pronoun resolution)~\citep{mathwin2023pronoun} requires resolving pronouns using gender information (e.g., ``The nurse sent the doctor a request because... [she]''). \texttt{copy} (Copy suppression)~\citep{mcdougall2023copy} investigates a negative mechanism where attention heads attend to previous token instances to suppress repetition. 
\texttt{induction} (Induction heads)~\citep{olsson2022induction} represent a foundational benchmark: given ``A B ... A'', predict ``B''. The circuit involves composition between Previous Token heads and Induction heads, demonstrating Q-K composition. \texttt{modular} (Modular addition)~\citep{nanda2023progress} requires reverse-engineering a transformer computing $(a + b) \mod p$. The circuit employs discrete Fourier transform features and trigonometric identities, testing whether agents can identify mathematical structure. \texttt{docstring} (Docstring completion)~\citep{heimersheim2023docstring} tests code understanding on a 4-layer attention-only transformer: given a function definition, the model predicts argument names in the docstring.  \texttt{balanced\_bracket} (Balanced bracket classification)~\citep{mcdougall2023brackets} classifies parenthesis sequences as balanced or unbalanced, testing whether agents can discover state-tracking mechanisms analogous to a counter. \texttt{max\_of\_k} (Max-of-K)~\citep{gross2024compact} requires predicting the maximum value from a list of integers. 

\textit{Open-ended Research Questions.} The open-ended tasks comprise ten exploratory research questions without established ground-truth circuits, testing whether agents can formulate coherent hypotheses and conduct rigorous investigations when outcomes are uncertain. These questions span linguistic phenomena, model limitations, and abstract reasoning capabilities.

\texttt{sarcasm} asks how sarcastic intent is represented, whether the model maintains separate representations for literal versus intended meaning, and how context modulates interpretation. \texttt{multilingual} explores how models handle instructions in different languages, probing whether internal processing is language-agnostic or maintains language-specific pathways. \texttt{typo}  investigates how models map misspelled tokens to intended forms, testing whether error correction uses explicit circuits or emerges from distributional robustness.
\texttt{unanswerable} probes how models recognize queries that cannot be answered from available context, testing for circuits that detect answerability. \texttt{uncertainty} asks whether dedicated circuits encode model confidence or epistemic uncertainty, distinct from prediction content. \texttt{count} investigates why models struggle to generate exactly $n$ items, exploring whether counting relies on approximate mechanisms that degrade with sequence length.
\texttt{inevitability} tests whether GPT-2 distinguishes semantic inevitability (physical causation, e.g., ``She dropped the glass. It hit the floor and...'') from narrative inevitability (story logic, e.g., ``The detective found the final clue. The mystery was...''). \texttt{irreversibility} probes which layers encode whether events can be undone (e.g., ``He closed the door'' vs. ``He broke the vase'').
\texttt{persona}~\cite{kozlov_persona_collapse} investigates why different assigned personas converge on identical preferences, probing whether persona representations are shallow or deeply integrated. \texttt{moral} asks whether models separate moral valence (whether an action is wrong) from consequentialist evaluation, testing for dissociable ethical representations.

\textit{Human-Written Repositories.} The human-written repositories comprise ten published research projects from the mechanistic interpretability literature, enabling evaluation of our pipeline on artifacts produced through standard academic workflows rather than agent generation. These repositories vary in scope, methodology, and documentation quality, providing a realistic test of generalization.

\texttt{filter}~\citep{sensharma2023filter} identifies attention heads that implement general list-filtering operations across diverse tasks. \texttt{universal} (Universal neurons)~\citep{gurnee2024universal} catalogs neurons in GPT-2 that activate consistently across contexts, proposing a taxonomy of interpretable features. \texttt{function\_vector}~\citep{todd2024function} demonstrates that in-context learning can be captured by single vectors that, when added to activations, induce task performance without exemplars.
\texttt{rome}~\citep{meng2022locating} locates factual associations in GPT and introduces Rank-One Model Editing for targeted knowledge modification. \texttt{relation} (Relation decoding)~\citep{hernandez2024linearity} shows that transformer language models decode relational knowledge through approximately linear maps from subject representations to object predictions. \texttt{erasing} (Concept erasing)~\citep{gandikota2024elm} proposes methods for removing specific concepts from language model representations, with applications to unlearning.
\texttt{belief} (Belief tracking)~\citep{prakash2025languagemodelsuselookbacks} investigates how language models represent characters' beliefs in theory-of-mind tasks, identifying ``lookback'' mechanisms that track belief states. \texttt{interpdetect}~\citep{tan2025interpdetect} develops interpretable signals for detecting hallucinations in retrieval-augmented generation for financial QA. \texttt{leela} (Chess reasoning)~\citep{sandmann2025iterativeinferencechessplayingneural} analyzes iterative inference in Leela Chess Zero, probing how neural networks refine position evaluations through computation.
\texttt{arithmetic} (Vector arithmetic)~\citep{feucht2025arithmetic} examines whether concept vectors support algebraic operations in both concept and token subspaces, testing compositionality of learned representations.

\para{Ablation Study Settings.} Doc-Only ablation removes access to all artifacts except the final documentation or report. The evaluator cannot see the research plan, code, or execution traces. All judgments are based solely on the written narrative. Consequently, it can assess only textual coherence and reported generalization, while reproducibility is inferred from descriptions rather than verified. Since some human-written repositories include code snippets in their documentation, we also include code evaluation. Specifically, we evaluate execution quality(\texttt{C1}--\texttt{C4}), consistency(\texttt{CS1}--\texttt{CS5}), generalizability(\texttt{GT1}--\texttt{GT3}), and replication quality(\texttt{RP1}--\texttt{RP3}). 

No-Execution ablation provides the evaluator with the full repository, including the plan, code, walkthrough, and report, but disallows code execution. Reproducibility and generalization are evaluated based solely on written evidence. Since instruction following and consistency evaluation do not involve code execution in our original pipeline, we exclude these metrics from this ablation. Specifically, we evaluate execution quality(\texttt{C1}--\texttt{C4}), generalizability(\texttt{GT1}--\texttt{GT3}), and replication quality(\texttt{RP1}--\texttt{RP3}). Comparing this ablation to our full pipeline isolates the value of execution-grounded evaluation, highlighting failures that cannot be detected through code inspection alone.

\begin{table}
  \caption{Structured binary checklist. Please refer to supplementary material for the complete description for each item.}
  \centering
  \begin{tabular}
  {p{0.17\textwidth}p{0.12\textwidth}p{0.65\textwidth}}
  
            \toprule
            Dimensions & Aspects & Checklists \\
            \midrule
            \textsc{Coherence} & Consistency Evaluation&\textbf{\texttt{CS1:}} All evaluable conclusions in the documentation match the results originally recorded in the notebook.\\
             && \textbf{\texttt{CS2:}} A plan file exists, and all steps in the final version of the plan are reflected in the implementation. \\
             && \textbf{\texttt{CS3:}} The reported effects have a clearly non-trivial magnitude (effect size) relative to baseline behavior or variability, such that the conclusions do not rely on marginal or negligible changes. \\
             && \textbf{\texttt{CS4:}} All key design choices and intermediate conclusions are explicitly justified, explaining why each design was chosen and how each conclusion is supported.  \\
             && \textbf{{\texttt{CS5:}}} Key experimental results supporting the main claims report appropriate measures of uncertainty or significance (e.g., error bars, confidence intervals, or statistical tests), with a clear explanation of what variability they capture. \\
             &Instruction Following& \textbf{\texttt{TS1:}} The goal described in the plan file matches the input stated goal. \\
             && \textbf{\texttt{TS2:}} The goal described in the plan file matches the input stated goal. \\
             && \textbf{\texttt{TS3:}} The plan file’s methodology follows the input intended direction and covers the required analyses. \\
             && \textbf{\texttt{TS4:}} For every circuit component identified, the tests confirm that its behavior matches the hypothesized function described in the given plan. \\ 
             
            \midrule
            \textsc{Reproducibility} &Execution Quality& \textbf{\texttt{C1:}} The block executes without error.  \\
            &&  \textbf{\texttt{C2:}} The logic implements the described computation correctly (indexing, metric formulas, patching logic, dataset handling). \\
            && \textbf{\texttt{C3:}} The block duplicates another block’s computation without adding new information. (Revising previous wrong results does not considered as redundant.) \\
            && \textbf{\texttt{C4:}} The block does not contribute to achieving the project goal as defined in the lan, code walkthrough, or documentation.\\
            &Replication Quality & \textbf{\texttt{RP1:}} The experiment can be reconstructed from the plan and code-walk without missing steps or required inference beyond ambiguous interpretation. \\
            && \textbf{\texttt{RP2:}} The environment (packages, models, data) can be restored and run without unresolved version or dependency issues. \\
            && \textbf{\texttt{RP3:}} Replicated results are stable across multiple runs.\\
            && \textbf{\texttt{RP4:}(only when demo exists)} pass when all of the following conditions are satisfied: (1) The demo can be executed or followed without referencing hidden or external materials. (2) Experiment or result claimed in the original paper / plan is can be demonstrated in the demo.
            (3) The demo specifies all required inputs, configurations, and execution steps needed to reproduce the demonstrated results. \\
            && \textbf{\texttt{DE1:}} Replicated documentation reports results (metrics, trends, qualitative findings) that match the original documentation within acceptable tolerance (within 5\% deviation). \\
            && \textbf{\texttt{DE2:}} The replicated documentation presents conclusions and interpretations consistent with the original. \\
            && \textbf{\texttt{DE3:}} No new information appears that is absent from or unsupported by the original documentation. \\
             \midrule
            \textsc{Generalizability} & Finding Generalizability& \textbf{\texttt{GT1:}} The newly-proposed concept is predictable on a new model, and can be verified through at least one example. \\
            & & \textbf{\texttt{GT2:}} The newly-proposed concept is predictable on a new data instance, and can be verified through at least one example. \\
            & Method Generalizability & \textbf{\texttt{GT3:}} If the work propose a new method, the method can be applied to another similar task, and can be verified through at lease one example. \\
            \bottomrule
\end{tabular}
\label{tab:checklist}
\end{table}
\begin{table}
  \caption{Task suite used to evaluate research agents across replication, open-ended questions, and human-authored research.}
  \centering
  \begin{tabular}
  {p{0.15\textwidth}p{0.8\textwidth}}
  
            \toprule
            Task Category &  Task Description \\
            \midrule
            \textsc{Replication} & \texttt{ioi:} Identify the circuit for predicting indirect objects in sentences~\cite{wang2022interpretability}. \\
            & \texttt{acronyms:} Find components for predicting multiple consecutive tokens in acronym completions~\cite{garcia2024does}.\\
            & \texttt{greater\_than:} Locate the circuit that determines if a year exceeds a previous year~\cite{hanna2023greaterthan}. \\
            & \texttt{docstring:} Find the circuit that predicts function argument names in Python docstrings~\cite{heimersheim2023docstring}. \\
            & \texttt{pronoun:} Identify how the model resolves pronouns to their referent entities~\cite{mathwin2023pronoun}. \\
            & \texttt{copy:} Investigate how copy suppression heads prevent repeated token predictions~\cite{mcdougall2023copy}. \\
            & \texttt{induction:} Explain the mechanism behind decreased probability for recently-seen tokens~\cite{olsson2022induction}. \\
            & \texttt{modular:} Reverse-engineer a model trained to compute modular addition~\cite{nanda2023progress}. \\
            & \texttt{balanced\_bracket:} Understand the circuit for classifying balanced brackets~\cite{mcdougall2023brackets}. \\
            & \texttt{max\_of\_k:} Locate the circuit that identifies the maximum value in a list~\cite{gross2024compact}. \\

            \midrule
            \textsc{Open-ended} & \texttt{sarcasm:} Identify how the model represents and detects sarcasm. \\
            & \texttt{multilingual:} Investigate how the model processes multilingual instructions. \\
            & \texttt{unanswerable:} Understand how the model handles unanswerable questions. \\
            & \texttt{uncertainty:} Determine whether a dedicated circuit exists for representing uncertainty. \\
            & \texttt{typo:} Explain how the model maps misspelled tokens to correct forms. \\
            & \texttt{persona:} Investigate why different personas converge to similar preferences~\cite{kozlov_persona_collapse}. \\
            & \texttt{inevitability:} Distinguish how the model represents semantic vs. narrative inevitability. \\
            & \texttt{irreversibility:} Find which layers encode the irreversibility of events. \\
            & \texttt{moral:} Separate the model's representations of moral wrongness from bad outcomes. \\
            & \texttt{count:} Understand why the model struggles to generate a specified number of words. \\
            
            \midrule
            \textsc{Human Repo} & \texttt{filter:} Identify general filter heads that process list elements~\cite{sensharma2023filter}. \\
            & \texttt{interpdetect:} Detect hallucination signals in financial question answering~\cite{tan2025interpdetect}. \\
            & \texttt{arithmetic:} Study vector arithmetic in concept and token subspaces~\cite{feucht2025arithmetic}. \\
            & \texttt{universal:} Identify neurons with consistent functions across contexts~\cite{gurnee2024universal}. \\
            & \texttt{leela:} Analyze iterative inference in a chess-playing neural network~\cite{sandmann2025iterativeinferencechessplayingneural}. \\
            & \texttt{relation:} Test linearity of relation decoding in transformer language models~\cite{hernandez2024linearity}. \\
            & \texttt{function\_vector:} Study how in-context learning is encoded as function vectors~\cite{todd2024function}. \\
            & \texttt{erasing:} Evaluate methods for removing conceptual knowledge from models~\cite{gandikota2024elm}. \\
            & \texttt{belief:} Investigate how models track beliefs using lookback mechanisms~\cite{prakash2025languagemodelsuselookbacks}. \\
            & \texttt{rome:} Locate and edit factual associations in GPT models~\cite{meng2022locating}. \\
            \bottomrule
\end{tabular}
\label{tab:tasklist}
\end{table}
\newpage
\section{Detailed Results}~\label{sec:appendix_results}

\begin{figure}[t]
    \centering
    \begin{subfigure}[t]{0.7\textwidth}
        \includegraphics[width=\textwidth]{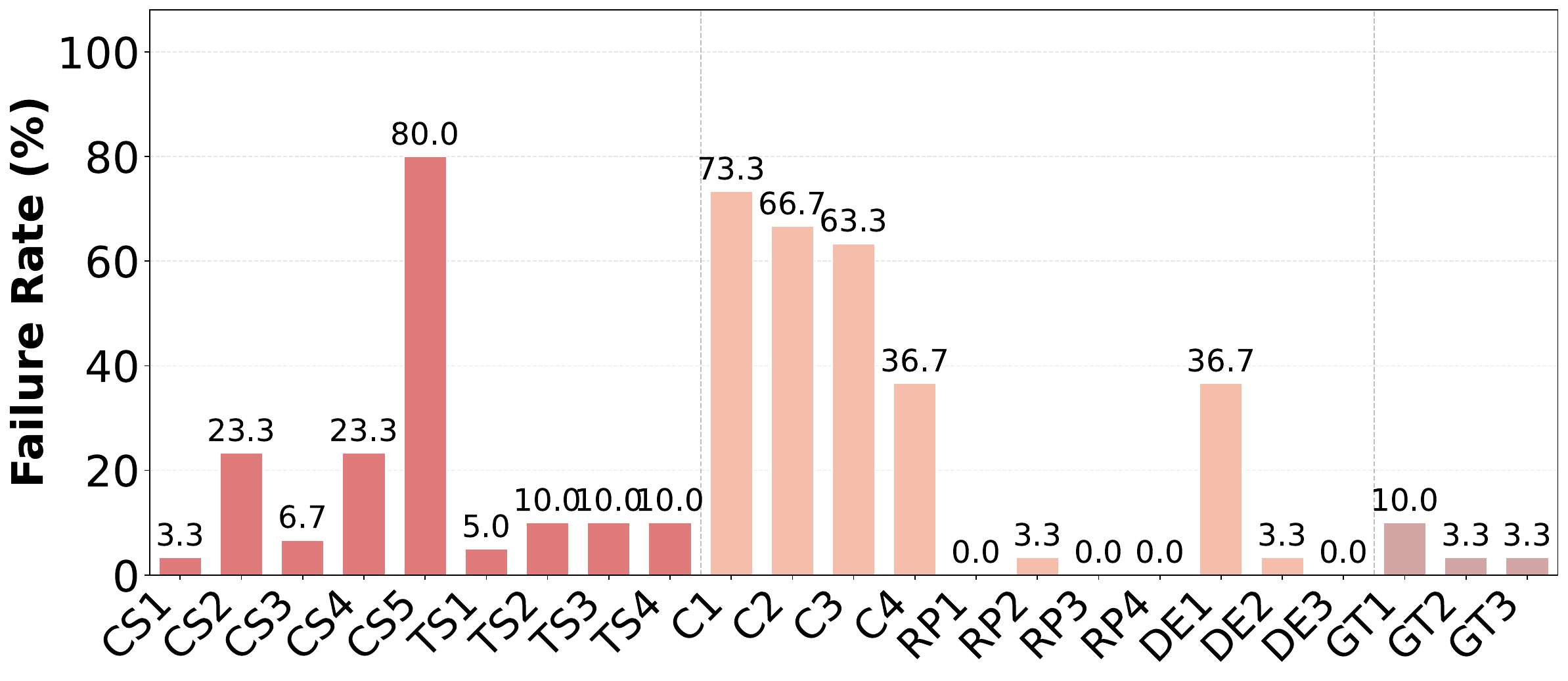}
        \caption{All Tasks}
        \label{subfig:failure_all_metrics}
    \end{subfigure}
    \begin{subfigure}[t]{0.7\textwidth}
        \includegraphics[width=\textwidth]{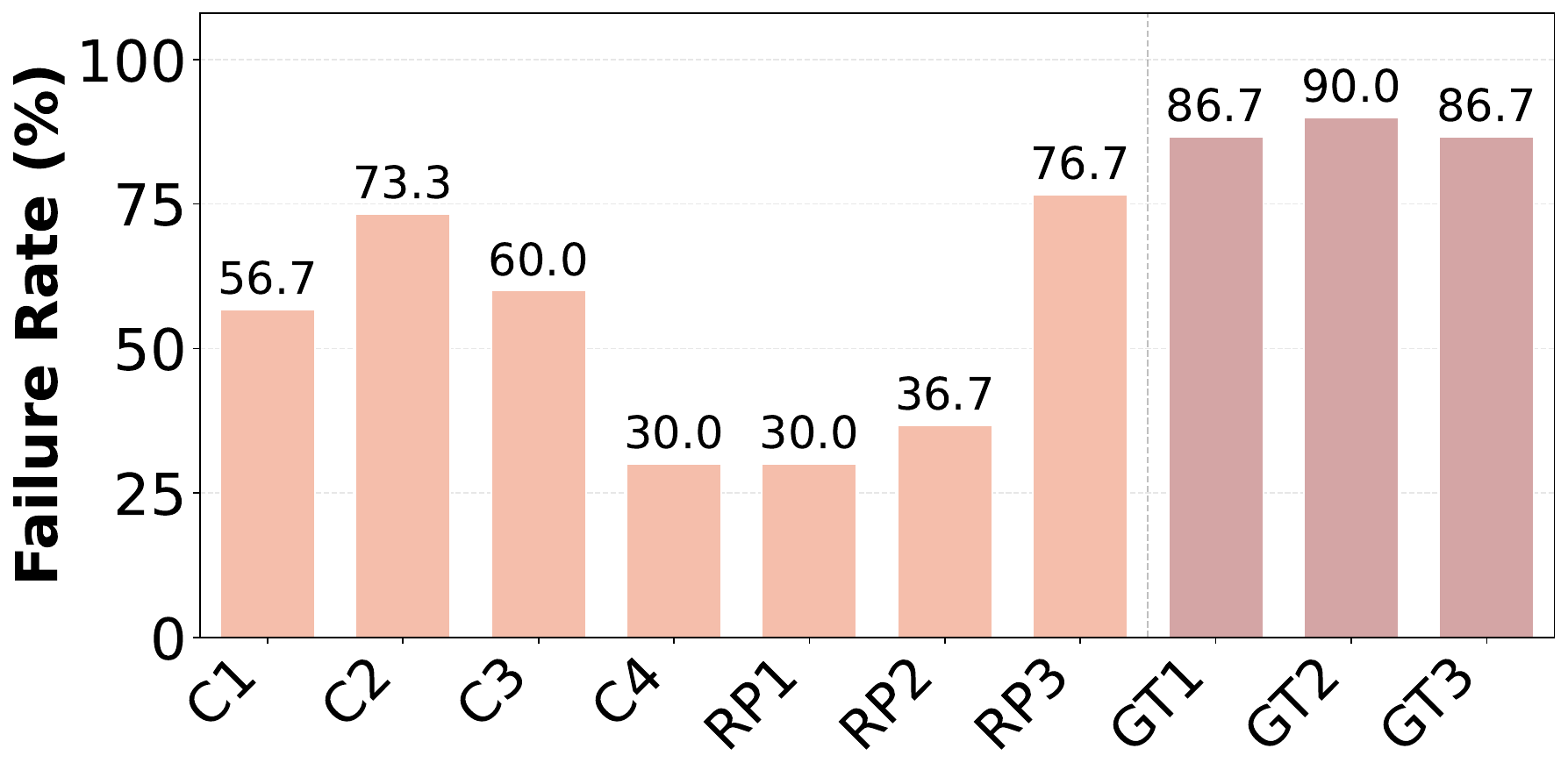}
        \caption{No Execution}
        \label{subfig:failure_all_metrics_no_exe}
    \end{subfigure}
    \begin{subfigure}[t]{0.7\textwidth}
        \includegraphics[width=\textwidth]{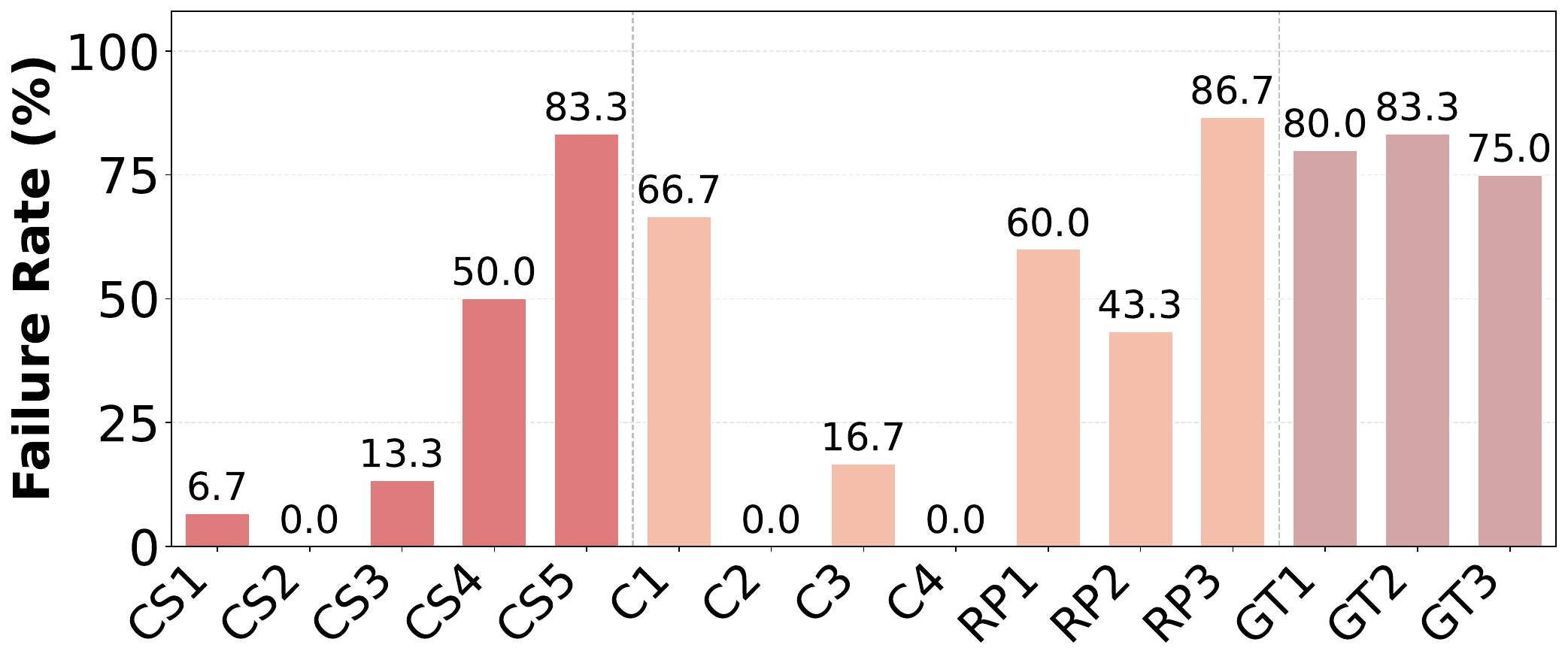}
        \caption{Doc Only}
        \label{subfig:failure_all_metrics_doc_only}
    \end{subfigure}
    \includegraphics[width=0.7\textwidth]{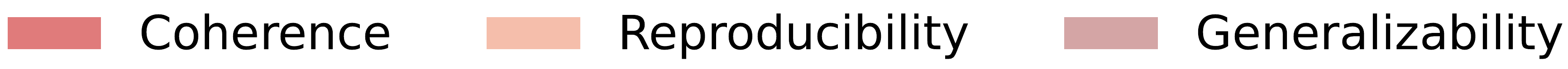}
    \caption{Average failure rate of all the tasks with \mechevalagent, Doc-only, and No-Execution. Under \mechevalagent evaluations, statistical significance(\texttt{CS5}) achieve 80\% failure rate. The failure rate in execution quality(\texttt{C1-C4}) and replicated results fidelity (\texttt{DE1}) are high. In contrast, FAIL is over-assigned in the Doc-only and No-Execution settings, especially in generalizability.}
    \label{fig:failure_all_metrics}
\end{figure}

\para{Detailed Failure Rates.}
Figure~\ref{fig:failure_all_metrics} shows the failure rate averaged across all tasks for each metric. \mechevalagent shows high failure rates in statistical significance and execution quality. The No-Execution and Doc-Only baselines exhibit even higher failure rates, especially in generalizability and replication quality where execution is crucial. However, agreement with human judgments is low for these baselines, indicating over-assignment of failures. 

\begin{figure}[!t]
    \centering
    \begin{subfigure}[t]{0.45\textwidth}
        \includegraphics[width=\textwidth]{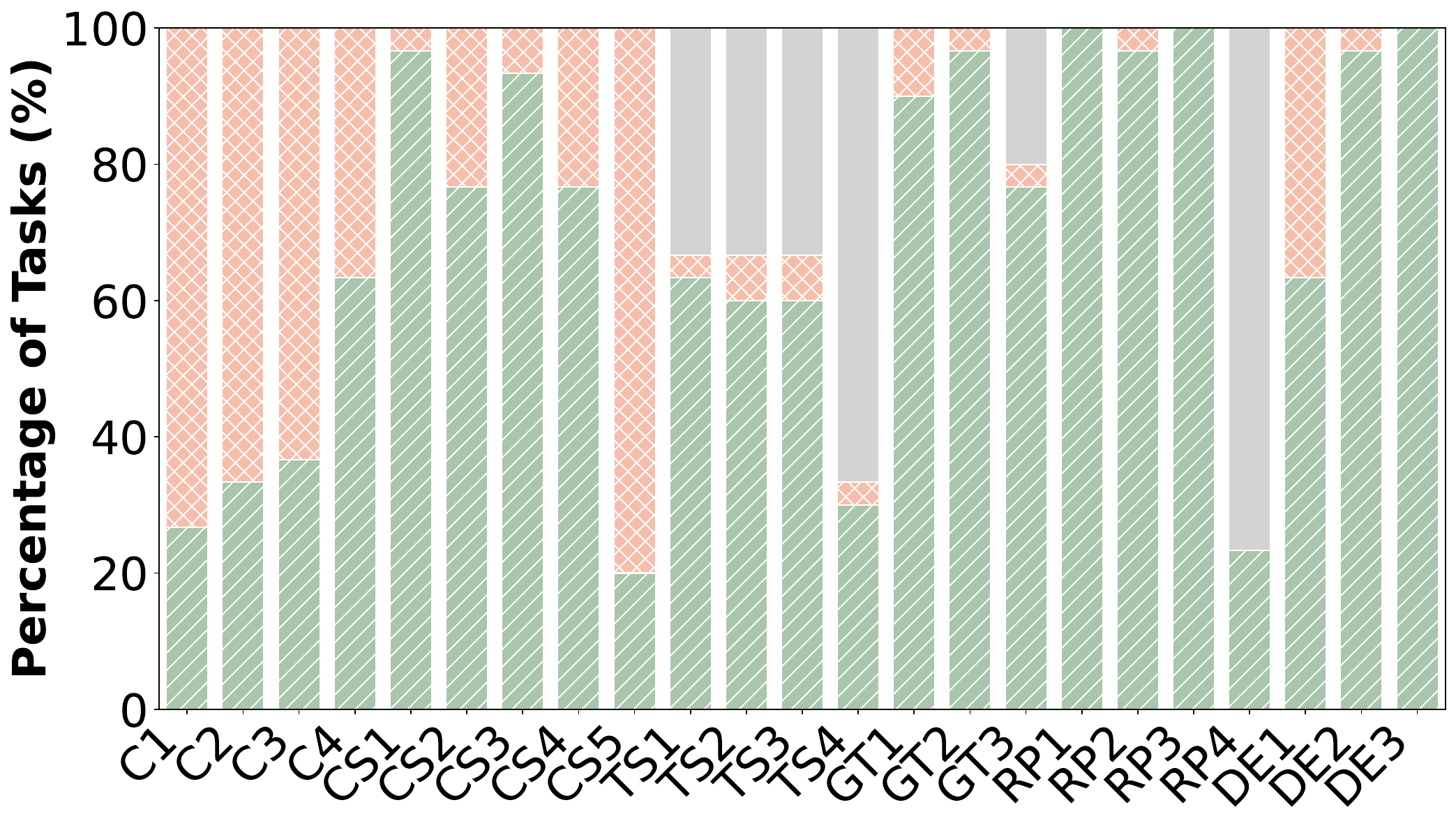}
        \caption{All Tasks}
        \label{subfig:eval_and_all}
    \end{subfigure}\hfill
    \begin{subfigure}[t]{0.45\textwidth}
        \includegraphics[width=\textwidth]{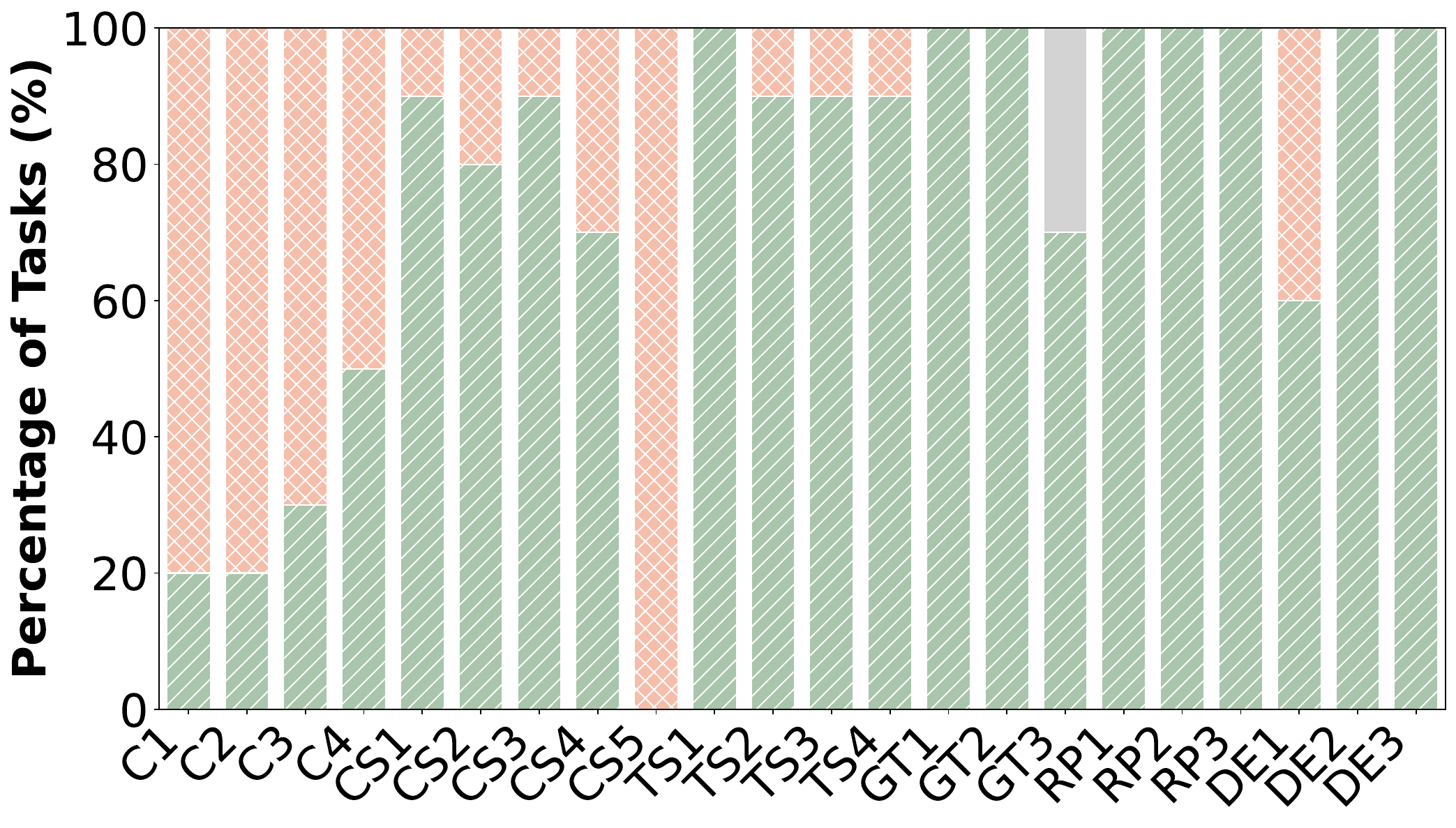}
        \caption{Replication}
        \label{subfig:eval_and_replication}
    \end{subfigure}

    \begin{subfigure}[t]{0.45\textwidth}
        \includegraphics[width=\textwidth]{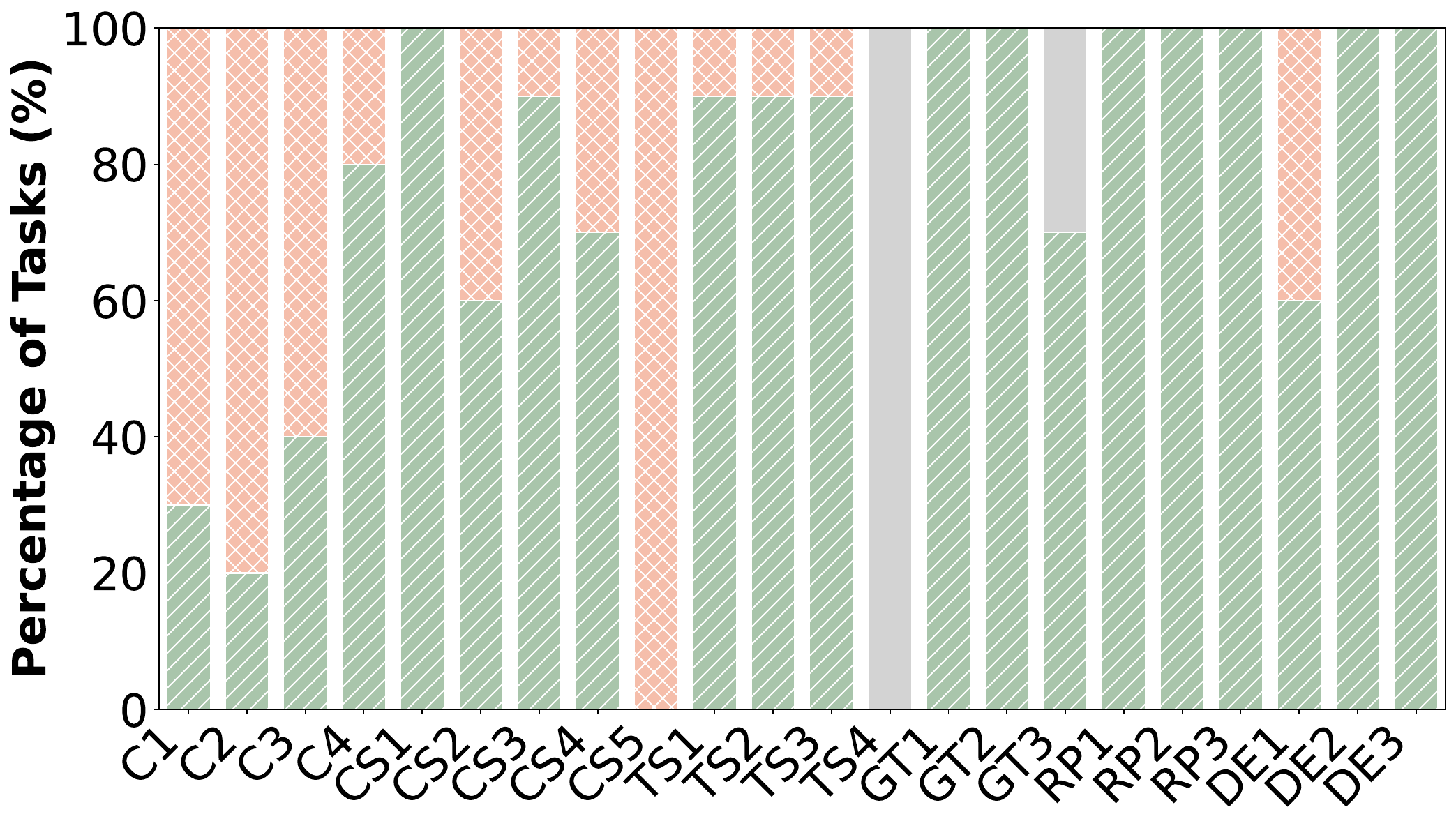}
        \caption{Open-ended}
        \label{subfig:eval_and_open}
    \end{subfigure}\hfill
    \begin{subfigure}[t]{0.45\textwidth}
        \includegraphics[width=\textwidth]{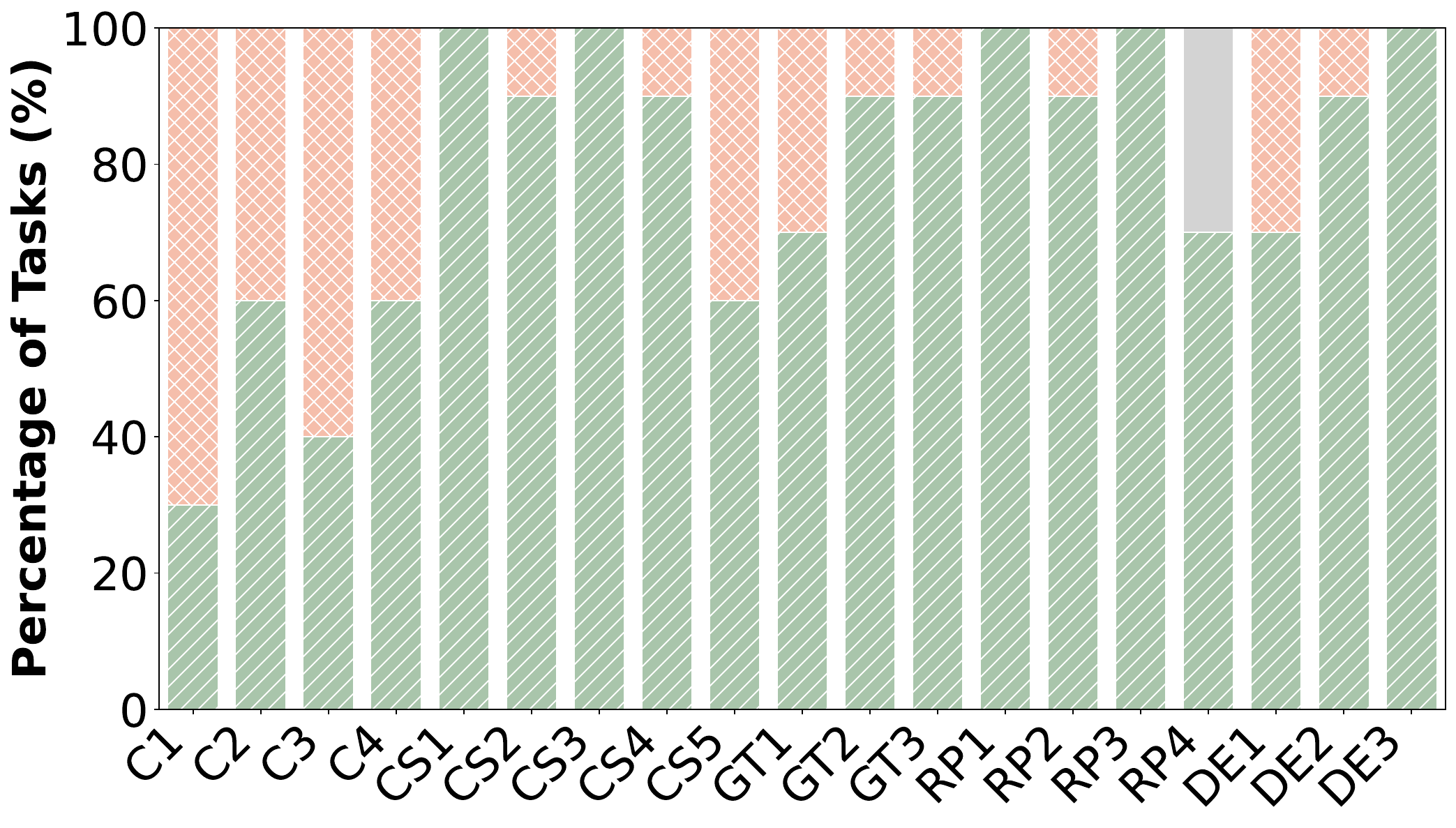}
        \caption{Human-written}
        \label{subfig:eval_and_human}
    \end{subfigure}

    \includegraphics[width=0.3\textwidth]{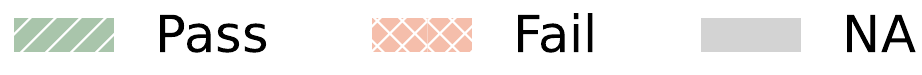}
    \caption{Pass rates using AND logic. (a) All 30 tasks combined. (b) Replication tasks. (c) Open-ended tasks. (d) Human-written repositories. Code quality metrics (C1-C4) and CS5 (result justification) show highest failure rates.}
    \label{fig:eval_and_logic_task}
\end{figure}

Figure~\ref{fig:eval_and_logic_task} shows pass rates for each checklist item broken down by task category. Overall pass rates are similar across categories. On average, human-written repositories achieve 78.9\%, replication tasks achieve 75.0\%, and open-ended tasks 75.7\%. Code quality metrics (\texttt{C1-C4}) show consistently high failure rates. Statistical significance (\texttt{CS5}) shows 100\% failure rate in both replication and open-ended tasks, while human-written repositories achieve 60\% pass rate. Replication tasks show relatively higher pass rates on instruction following (\texttt{TS1-TS4}), as expected given their well-defined experimental goals. Open-ended tasks show more variation in coherence metrics, reflecting the greater ambiguity in evaluating exploratory research. 

\begin{figure}[!t]
    \centering
    \begin{subfigure}[t]{0.43\textwidth}
        \includegraphics[width=\textwidth]{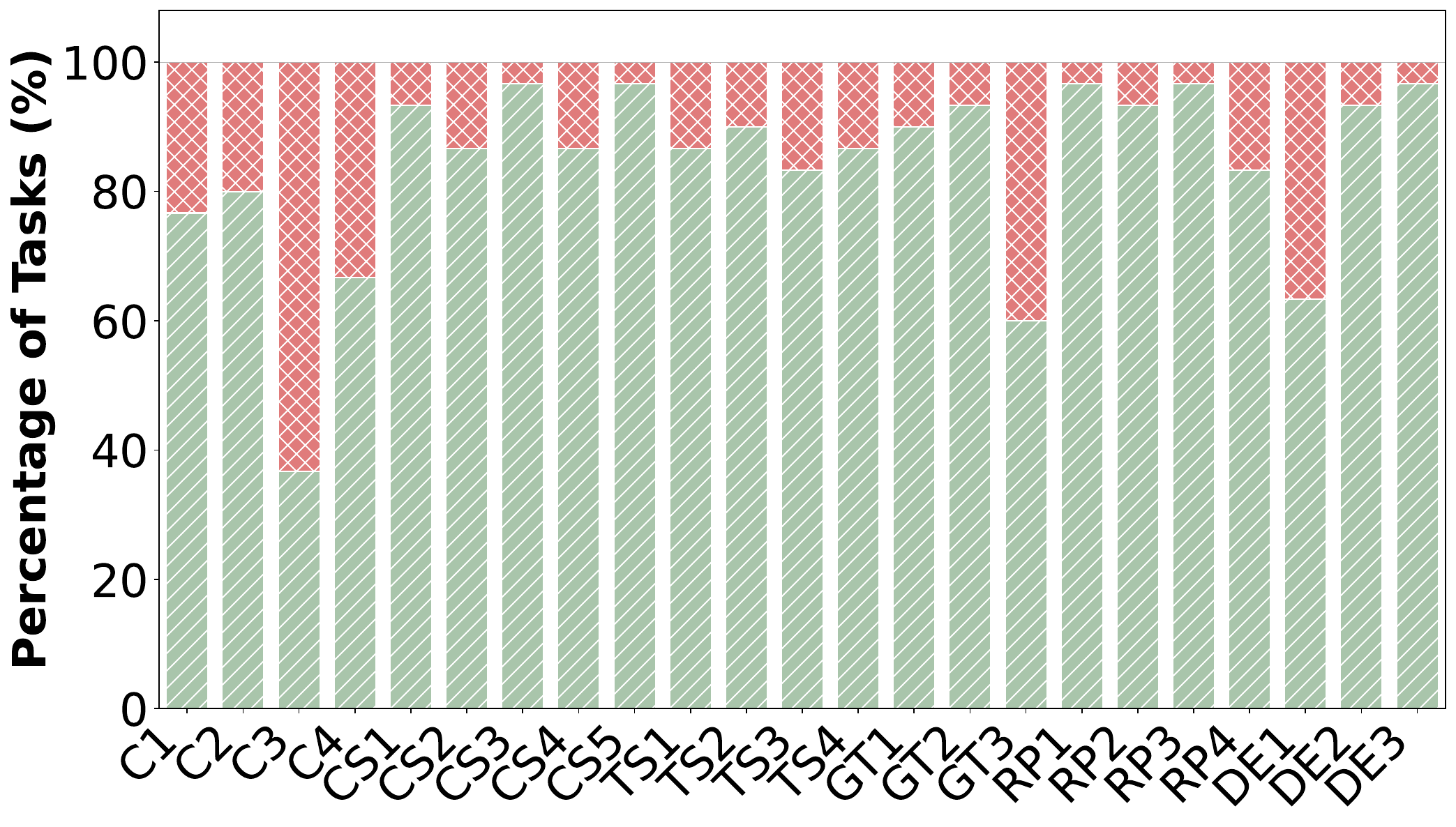}
        \caption{All Tasks}
        \label{subfig:and_vs_all}
    \end{subfigure}\hfill
    \begin{subfigure}[t]{0.43\textwidth}
        \includegraphics[width=\textwidth]{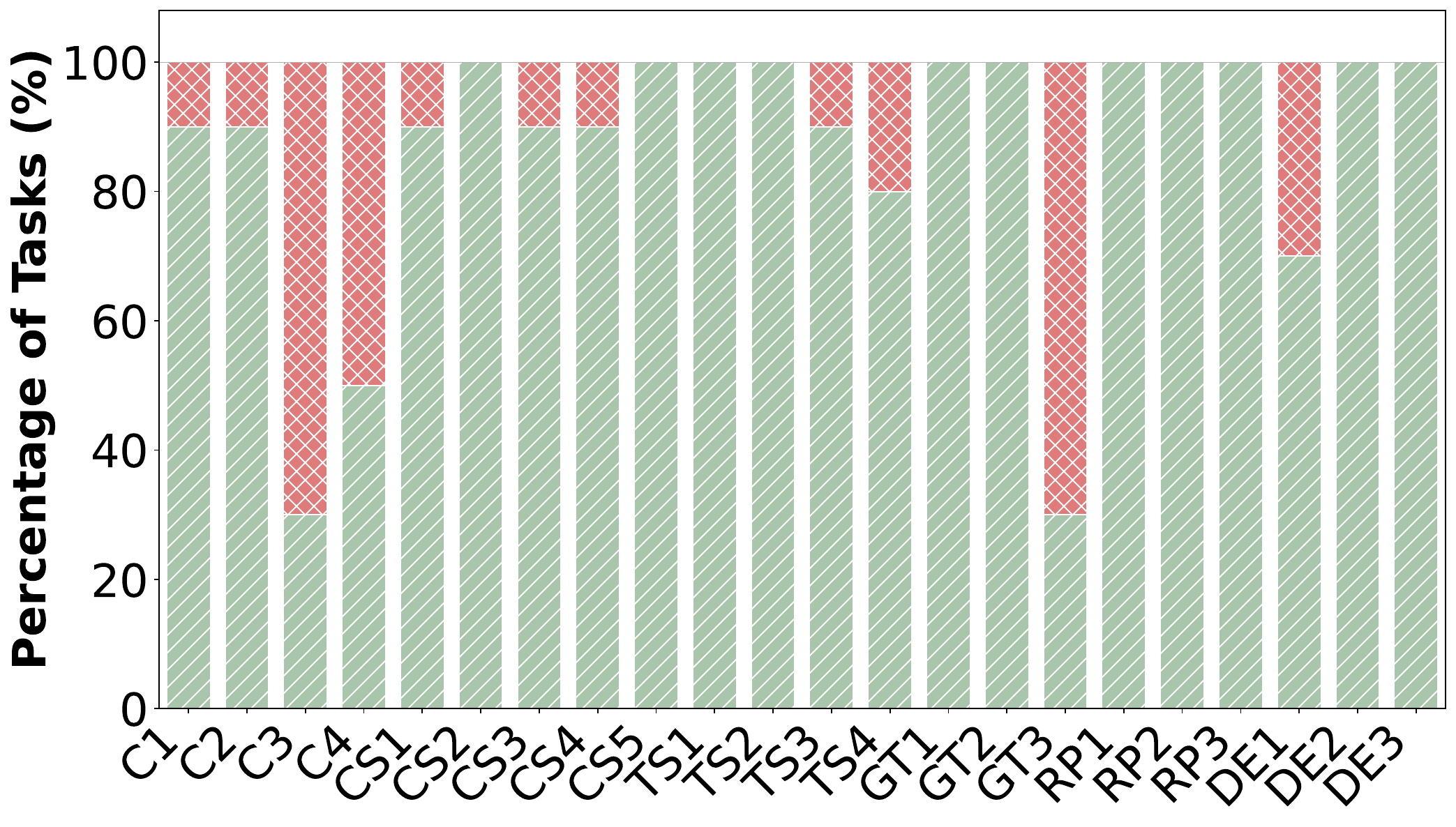}
        \caption{Replication}
        \label{subfig:and_vs_replication}
    \end{subfigure}

    \begin{subfigure}[t]{0.43\textwidth}
        \includegraphics[width=\textwidth]{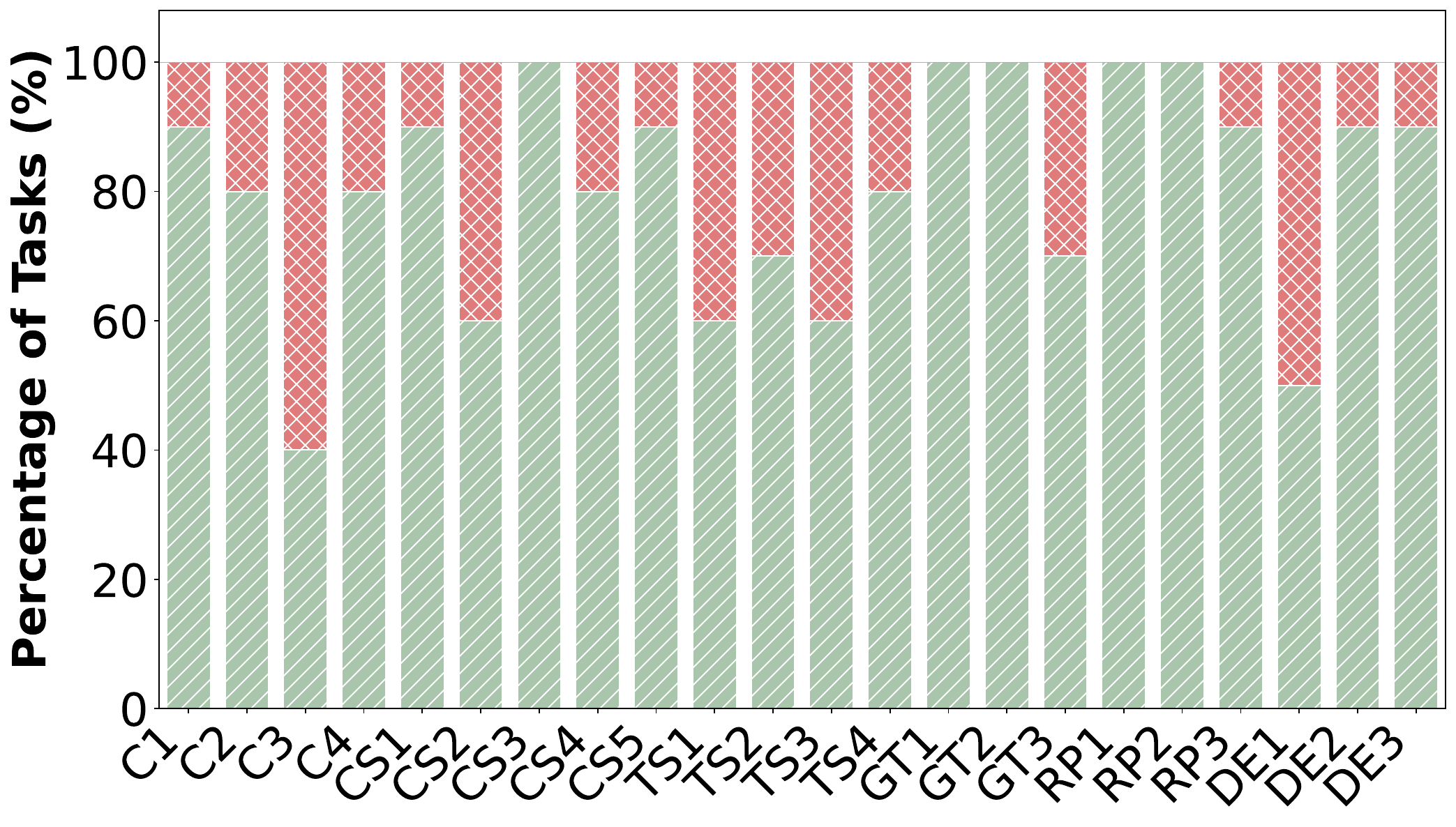}
        \caption{Open-ended}
        \label{subfig:and_vs_open}
    \end{subfigure}\hfill
    \begin{subfigure}[t]{0.43\textwidth}
        \includegraphics[width=\textwidth]{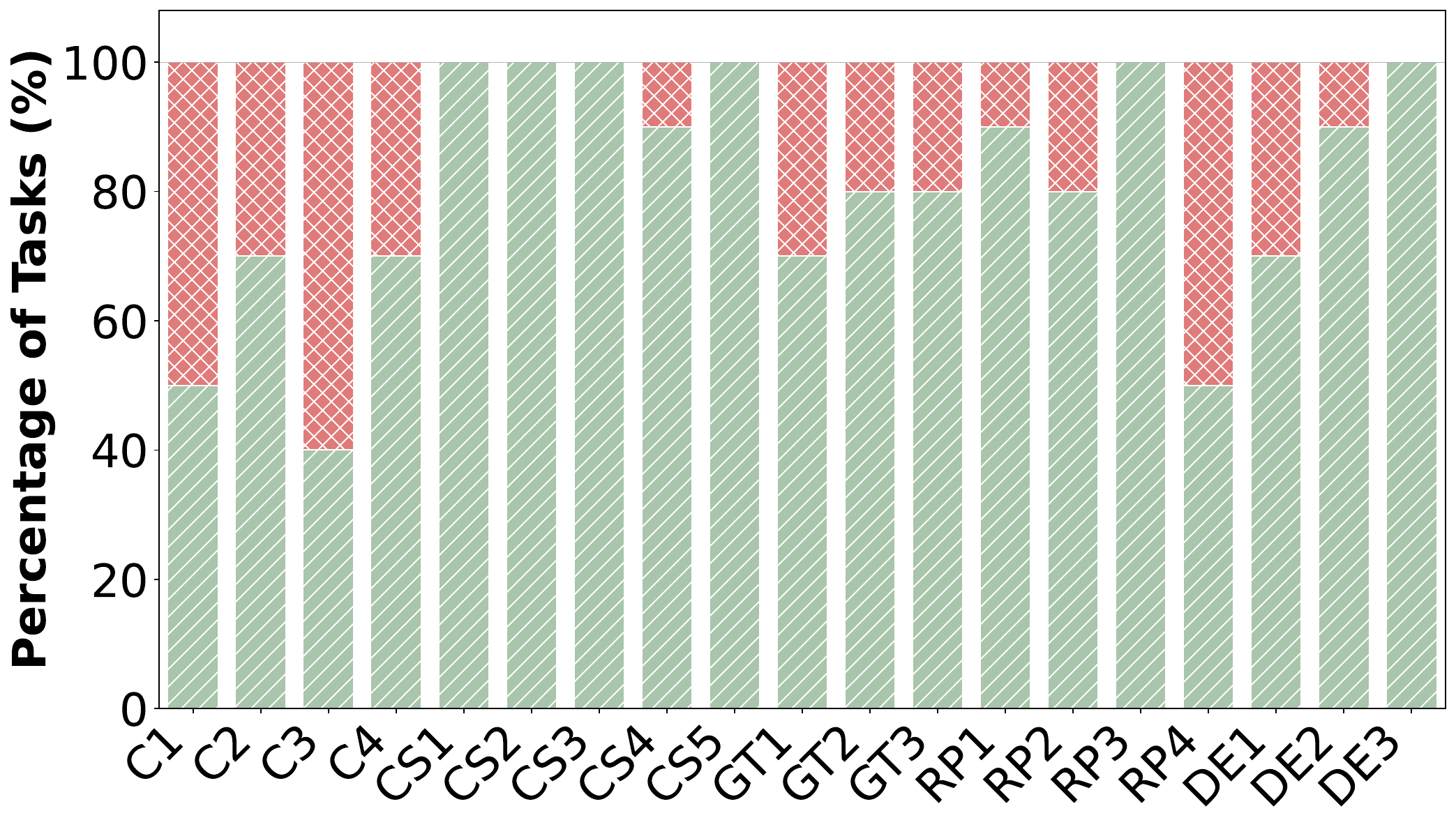}
        \caption{Human-written}
        \label{subfig:and_vs_human}
    \end{subfigure}

    \includegraphics[width=0.3\textwidth]{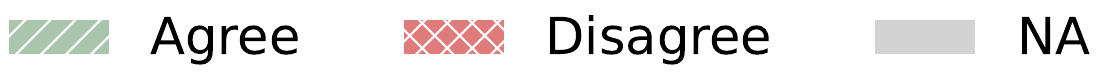}
    \caption{Agreement between \mechevalagent and human experts. Bars show percentage of tasks where agent's evaluation matched human assessment. (a) All tasks (above 80\% average). (b) Replication tasks. (c) Open-ended tasks. (d) Human-written repositories.}
    \label{fig:and_vs_human_task}
\end{figure}

\begin{figure}[t]
    \centering
    \includegraphics[width=0.7\textwidth]{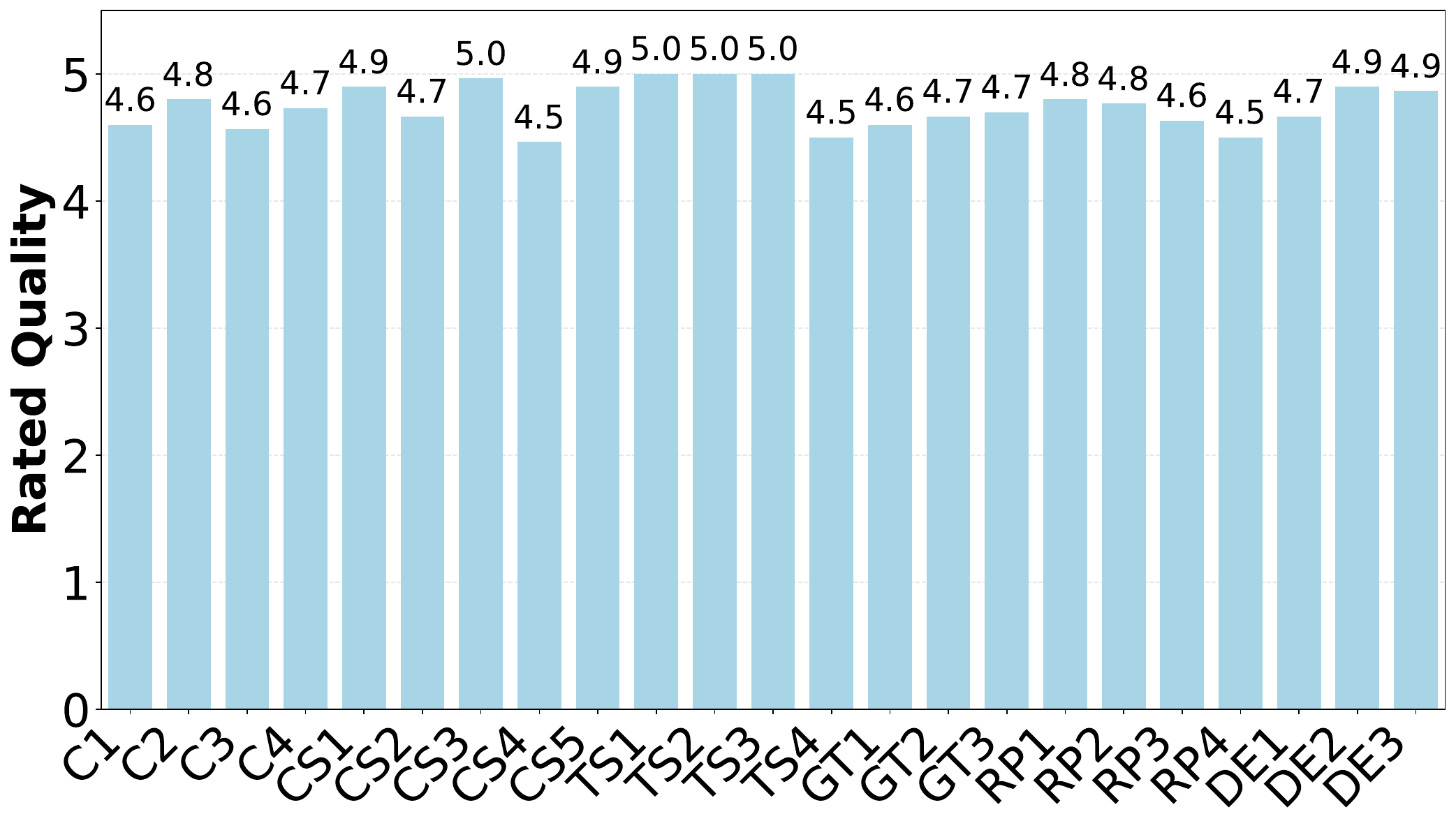}
    \caption{Human-rated quality on \mechevalagent evaluations (1-5 Likert scale) on each metrics averaged across three types of tasks.}
    \label{fig:mean_rating}
\end{figure}

\para{Agreement and Rated Quality.}
We measure whether \mechevalagent verdicts match human evaluations using the same checklist (agreement), and  how much human experts agree with the rationales \mechevalagent provides (rated quality). As shown in Figure~\ref{fig:and_vs_human_task}, \mechevalagent achieves high agreement with human experts across all task categories: 86.4\% for replication tasks, 80.5\% for open-ended tasks, and 80.5\% for human-written repositories. Consistency metrics (\texttt{CS1-CS5}) show particularly strong agreement, reaching 90-100\% across most categories. Reproducibility metrics (\texttt{RP1-RP3}) also achieve high agreement rates of 90-100\%. The lowest agreement occurs in \texttt{C3} (Redundancy), with only 30-40\% agreement across categories, and \texttt{GT3} (method generalizability), which shows 30\% agreement in replication tasks. Those failures are likely due to different interpretation of the checklist as we discussed in Section~\ref{subsec:closer_failure}. 
Figure~\ref{fig:mean_rating} shows average rated quality of \mechevalagent evaluations, which is above 4 (Agree) across all metrics. 

\begin{figure}[t]
    \centering
        \includegraphics[width=0.7\textwidth]{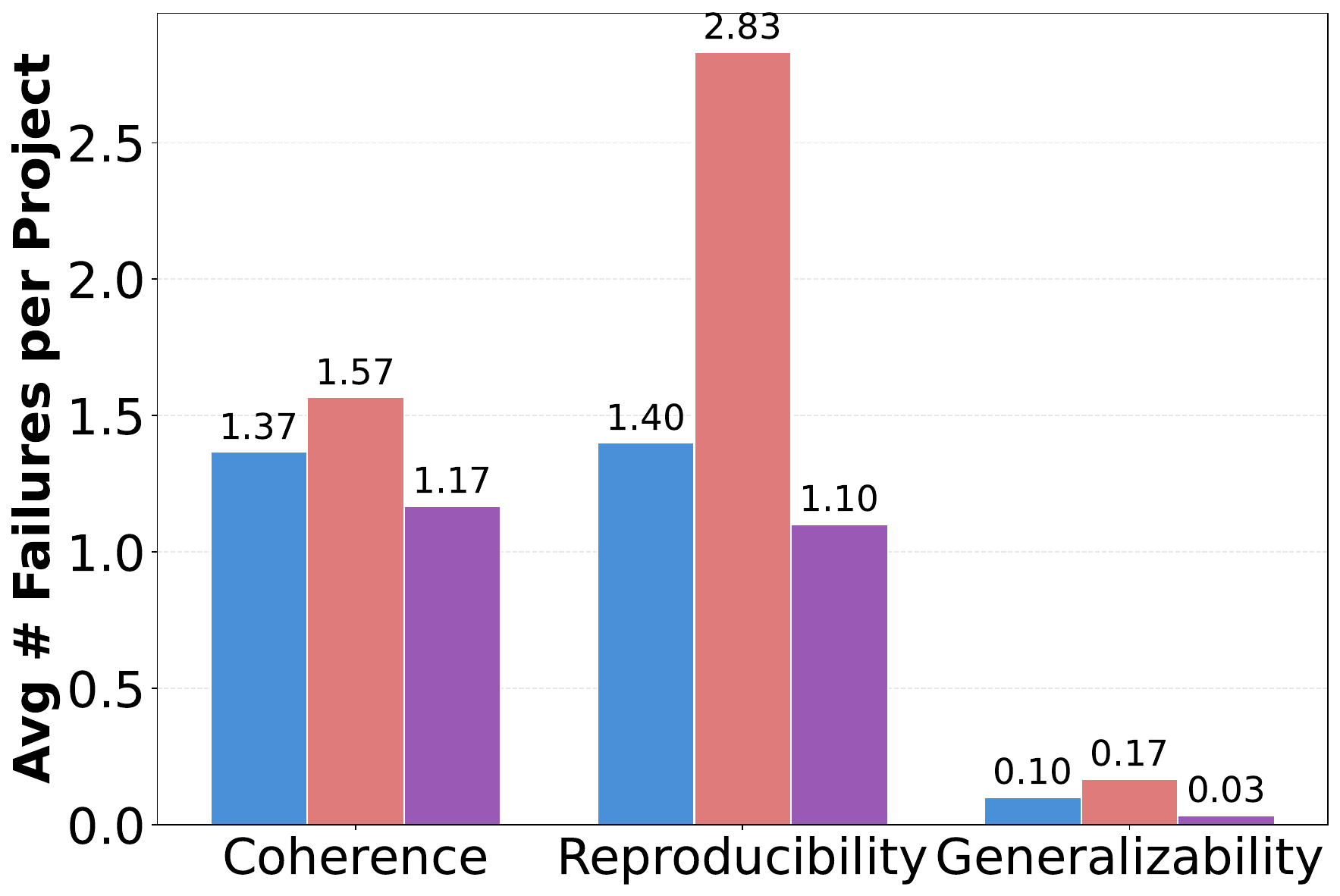}
        
        \includegraphics[width=0.6\textwidth]{plots/final/failure_breakdown_legend.pdf}
    \caption{Average number of unique issues identified by \mechevalagent per project, broken down by evaluation dimension. Reproducibility shows the most unique issues (2.83 per project), followed by coherence (1.57) and generalizability (0.17).}
    \label{fig:failure_number}
\end{figure}

Beyond agreement with human evaluation, \mechevalagent also identifies unique issues that humans miss, as discussed in Section~\ref{subsec:closer_failure}. On average, each project has 1.57 issues in coherence, 2.83 issues in reproducibility, and 0.17 issues in generalizability that are uniquely identified by \mechevalagent, as shown in Figure~\ref{fig:failure_number}. Reproducibility is the dimension where most unique issues are found. Fewer issues are found in generalizability, primarily due to our task design. As shown in Figure~\ref{fig:eval_and_logic_task}, all generalizability failures occur in human-written repositories. This is partially because our replication and open-ended tasks are not designed for specific architectures, whereas human-written repositories often contain demonstration code tailored to particular architectures. Additionally, passing the generalizability checklist does not guarantee that results transfer to other models or that the methodology is stable across settings, as discussed in Section~\ref{subsec:closer_failure}.

\para{Efficiency Comparison.}
Table~\ref{tab:efficiency} shows the average evaluation time per task across different categories. Replication tasks require the shortest evaluation time (1.0 hour on average), as they have well-defined experimental setups and smaller codebases. Open-ended questions take longer (2.3 hours) due to the need to understand the hypothesis and method proposed by research agents. Human-written repositories require the most time (3.35 hours), reflecting their larger codebases, more complex documentation, and the additional effort needed to extract goals and methodology from accompanying papers.
\begin{table}[t]
    \centering
    \caption{Average evaluation time per task by category. Human-written repositories require the longest evaluation time due to their larger codebases and more complex documentation.}
    \label{tab:efficiency}
    \begin{tabular}{lc}
        \toprule
        Task Category & Average Time \\
        \midrule
        Replication Tasks & 1.0 hr \\
        Open-ended Questions & 2.3 hr \\
        Human-written Repositories & 3.35 hr \\
        \bottomrule
    \end{tabular}
\end{table}

\begin{figure}[t]
    \centering
    \begin{subfigure}[t]{0.43\textwidth}
        \includegraphics[width=\textwidth]{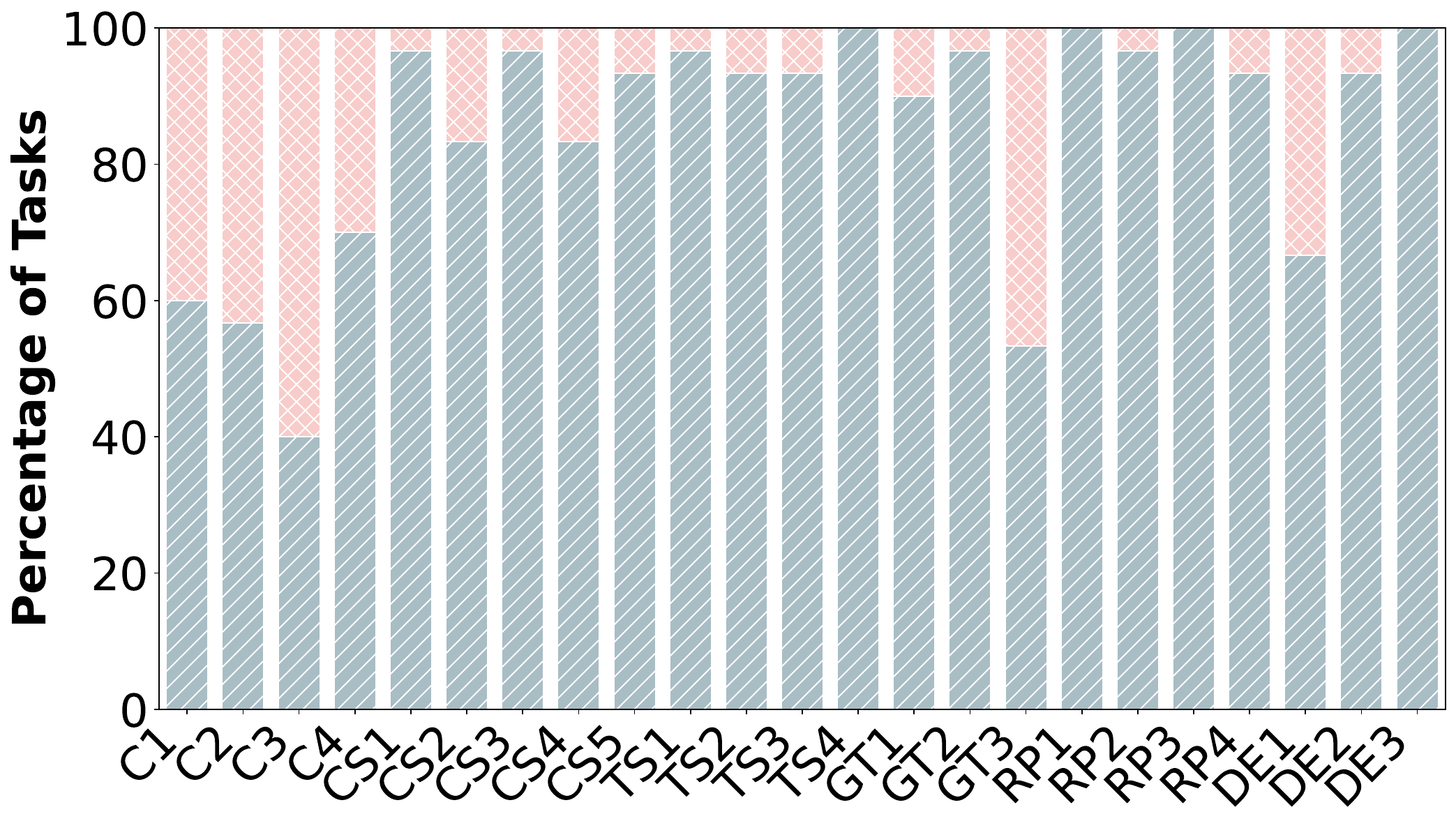}
        \caption{All Tasks}
        \label{subfig:metric_agreement_all}
    \end{subfigure}\hfill
    \begin{subfigure}[t]{0.43\textwidth}
        \includegraphics[width=\textwidth]{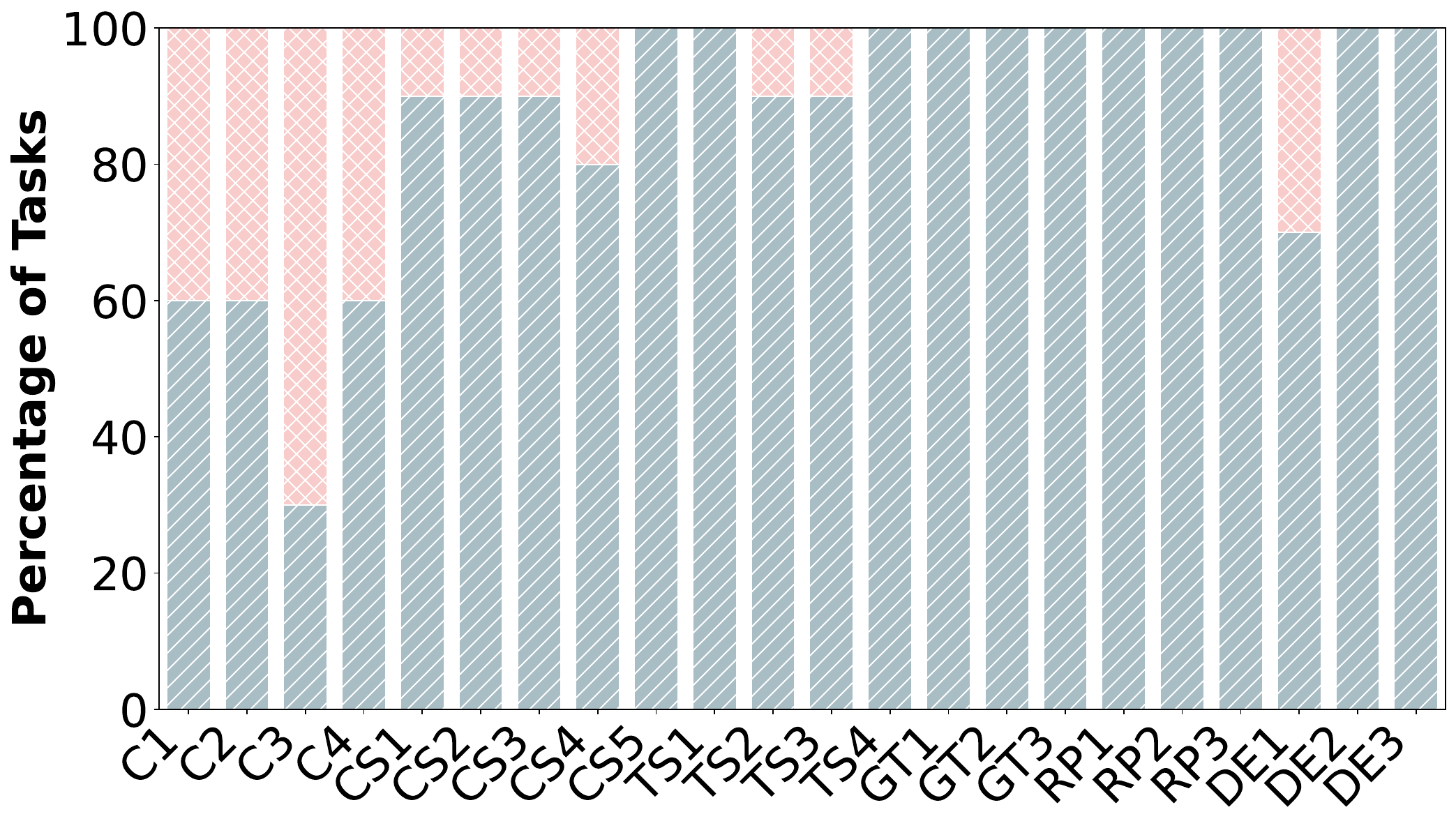}
        \caption{Replication}
        \label{subfig:metric_agreement_replication}
    \end{subfigure}

    \begin{subfigure}[t]{0.43\textwidth}
        \includegraphics[width=\textwidth]{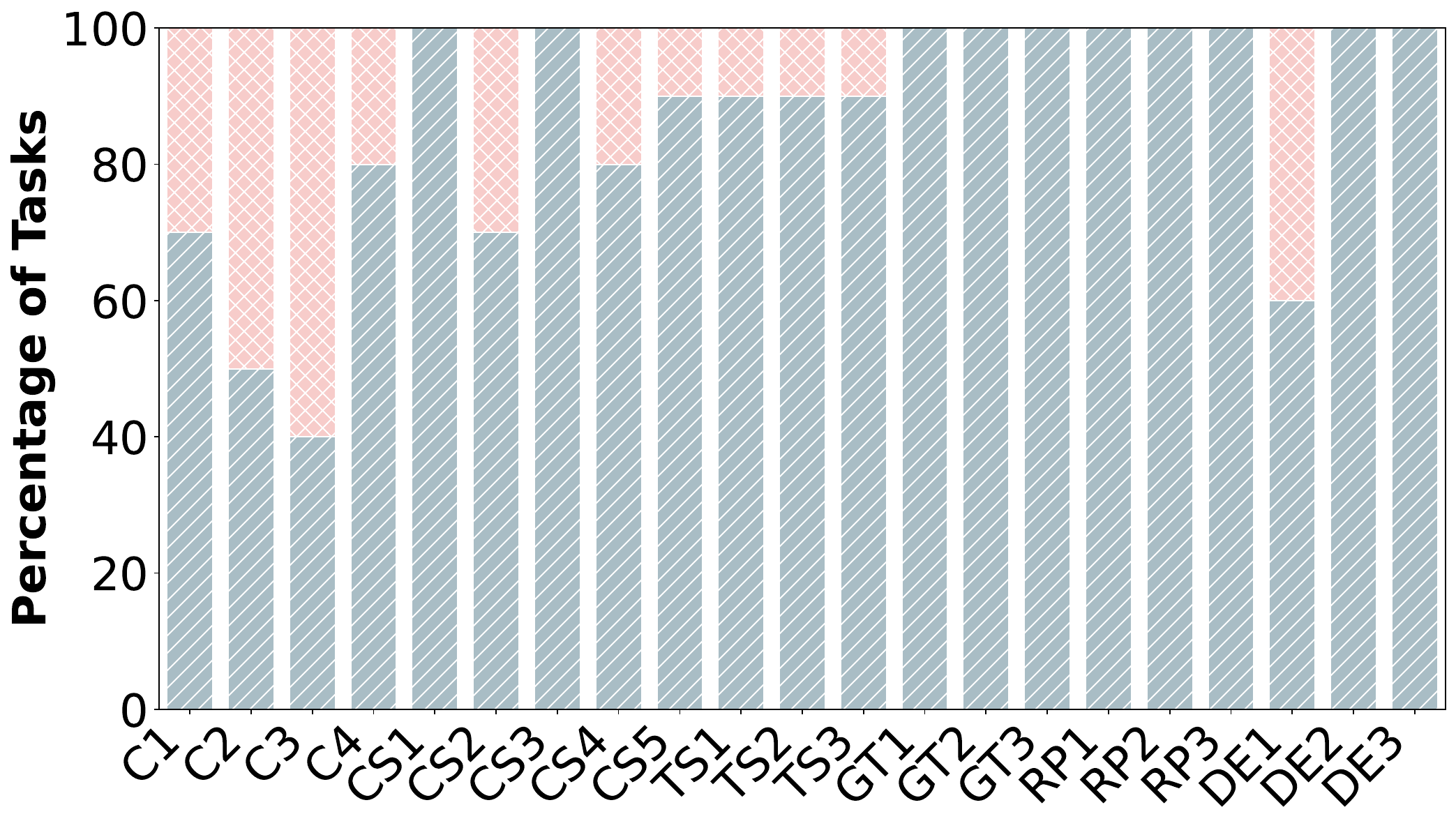}
        \caption{Open-ended}
        \label{subfig:metric_agreement_open}
    \end{subfigure}\hfill
    \begin{subfigure}[t]{0.43\textwidth}
        \includegraphics[width=\textwidth]{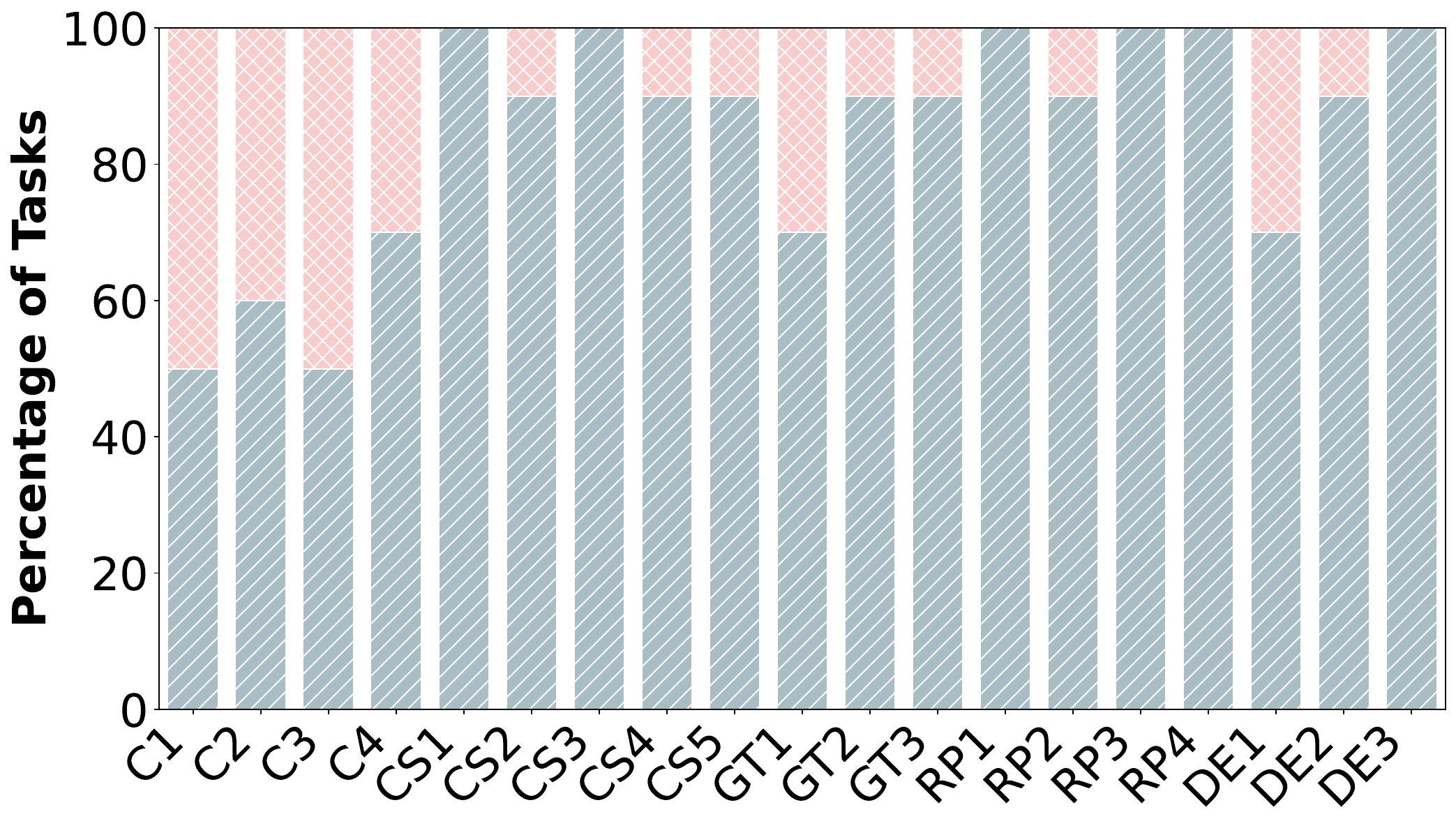}
        \caption{Human-written}
        \label{subfig:metric_agreement_human}
    \end{subfigure}

    \includegraphics[width=0.35\textwidth]{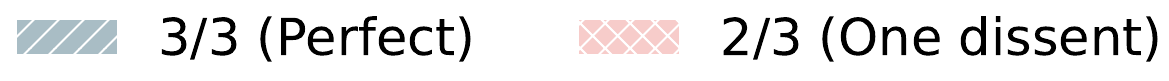}
    \caption{Evaluation stability across three runs. Bars show 3/3 (perfect) vs 2/3 (one dissent) agreement. (a) All tasks combined. (b) Replication tasks. (c) Open-ended tasks. (d) Human-written repositories. Most metrics achieve over 90\% perfect agreement.}
    \label{fig:metric_agreement_task}
\end{figure}

\begin{figure}[t]
    \centering
    \begin{subfigure}[t]{0.32\textwidth}
        \includegraphics[width=\textwidth]{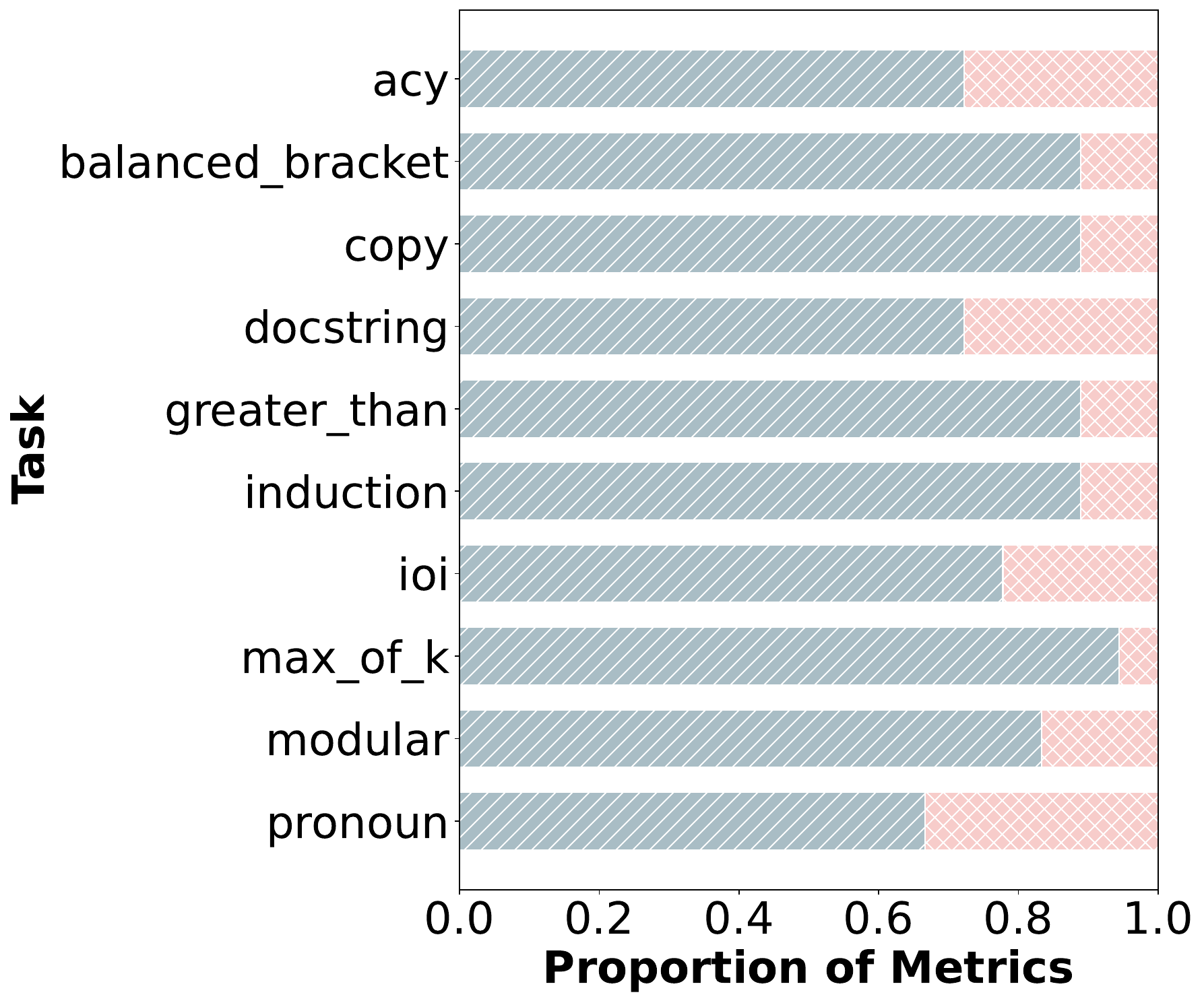}
        \caption{Replication}
        \label{subfig:stability_replication}
    \end{subfigure}\hfill
    \begin{subfigure}[t]{0.32\textwidth}
        \includegraphics[width=\textwidth]{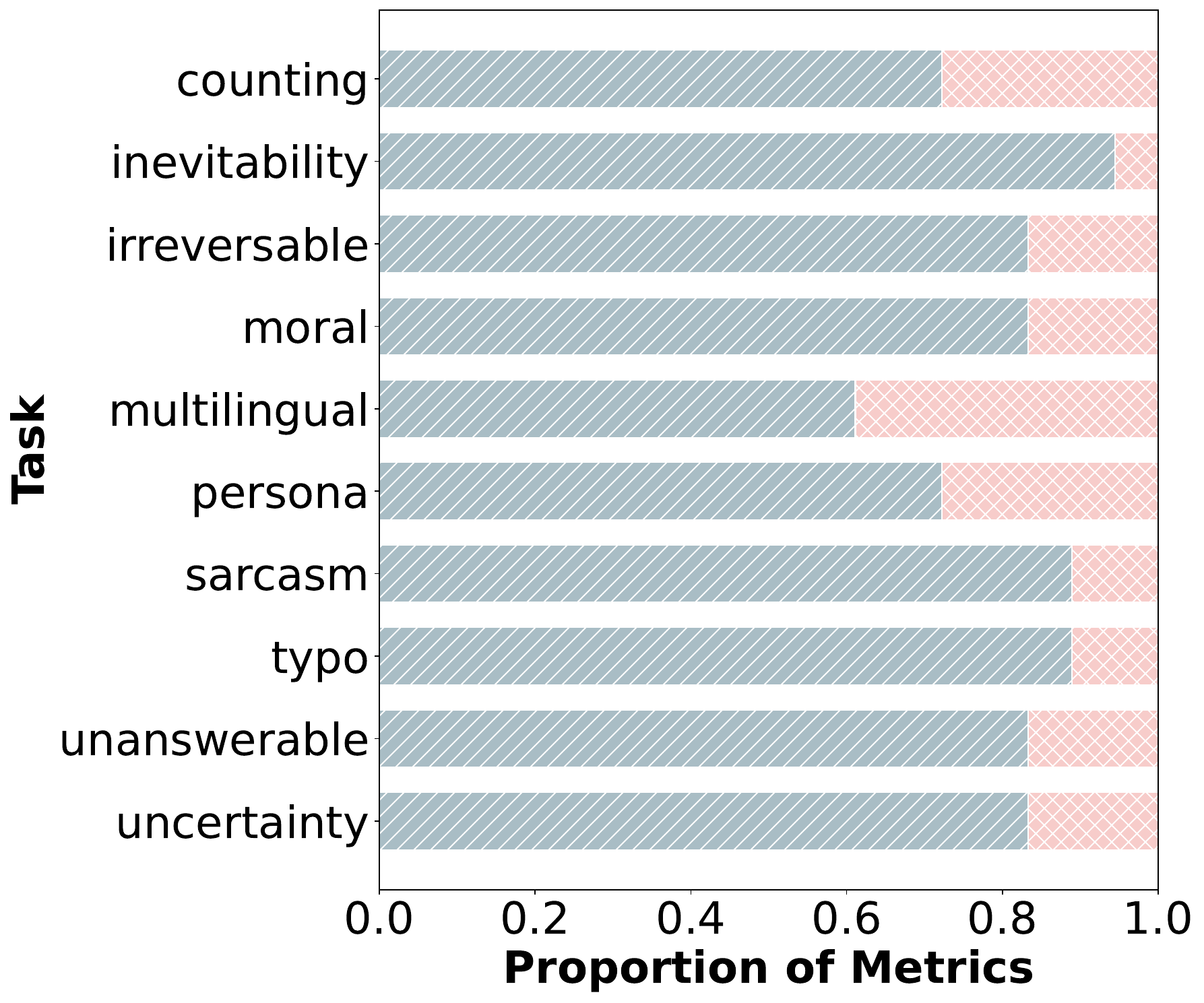}
        \caption{Open-ended}
        \label{subfig:stability_open}
    \end{subfigure}\hfill
    \begin{subfigure}[t]{0.32\textwidth}
        \includegraphics[width=\textwidth]{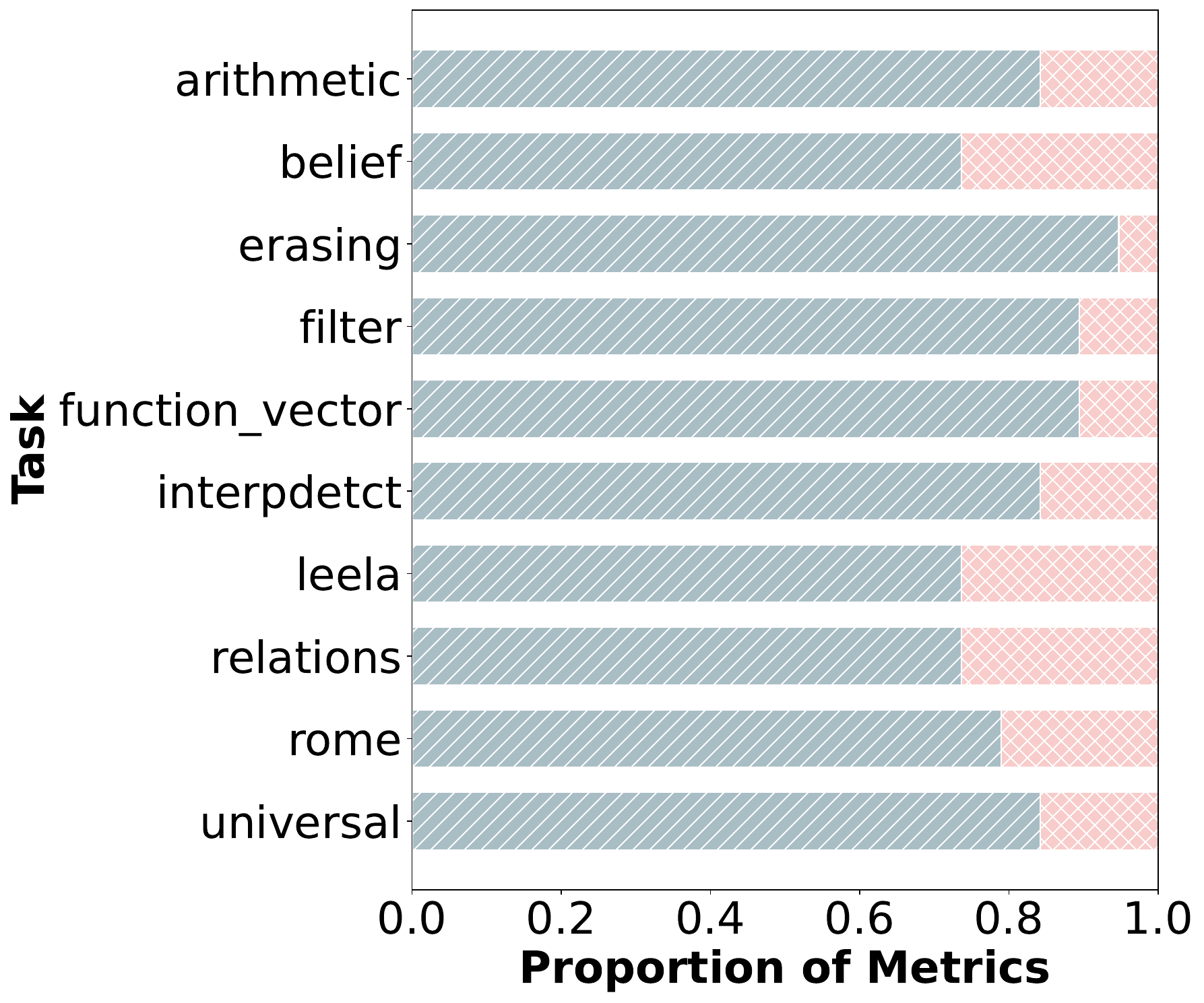}
        \caption{Human agreement}
        \label{subfig:stability_human}
    \end{subfigure}
    \includegraphics[width=0.35\textwidth]{plots/agreement_composition_legend.pdf}
    \caption{Evaluation stability by repository type. Bars show 3/3 (perfect) vs 2/3 (one dissent) agreement across three runs. (a) Replication tasks. (b) Open-ended tasks. (c) Human-written repositories.}
    \label{fig:stability_task}
\end{figure}

\para{Stability.}
We run evaluation for three times for each research project. Here, we measure the stability by checking the agreement across three runs. In Figure~\ref{fig:metric_agreement_task} and Figure~\ref{fig:stability_task}, we show the proportion of perfect agreement (3 runs are the same) and one-dissent agreement (1 run is different).

Our evaluations in consistency, instruction following, generalizability, and reproduction are mostly stable, as shown in Figure~\ref{subfig:metric_agreement_all}. We attribute the instability to two main factors. First, our binary checklists cannot enumerate every possible situation, particularly given the diverse input space of research artifacts, leading to inconsistent interpretations across runs. Second, we observe an interesting behavioral pattern in Claude Code, which is a reluctance to provide negative feedback. This can contribute to the instability during evaluation. The evaluation agent tends to exploit ambiguities in the checklist to assign PASS, which we discuss further in Section~\ref{sec:discuss}. As discussed in Section~\ref{subsec:closer_failure}, different interpretations of the checklist also contribute to the instability. Despite these sources of instability, agreement with human evaluators remains high, and the instability itself serves as a useful signal, highlighting cases that warrant closer human inspection.  

\begin{figure}[t]
    \centering
    \begin{subfigure}[t]{0.48\textwidth}
        \includegraphics[width=\textwidth]{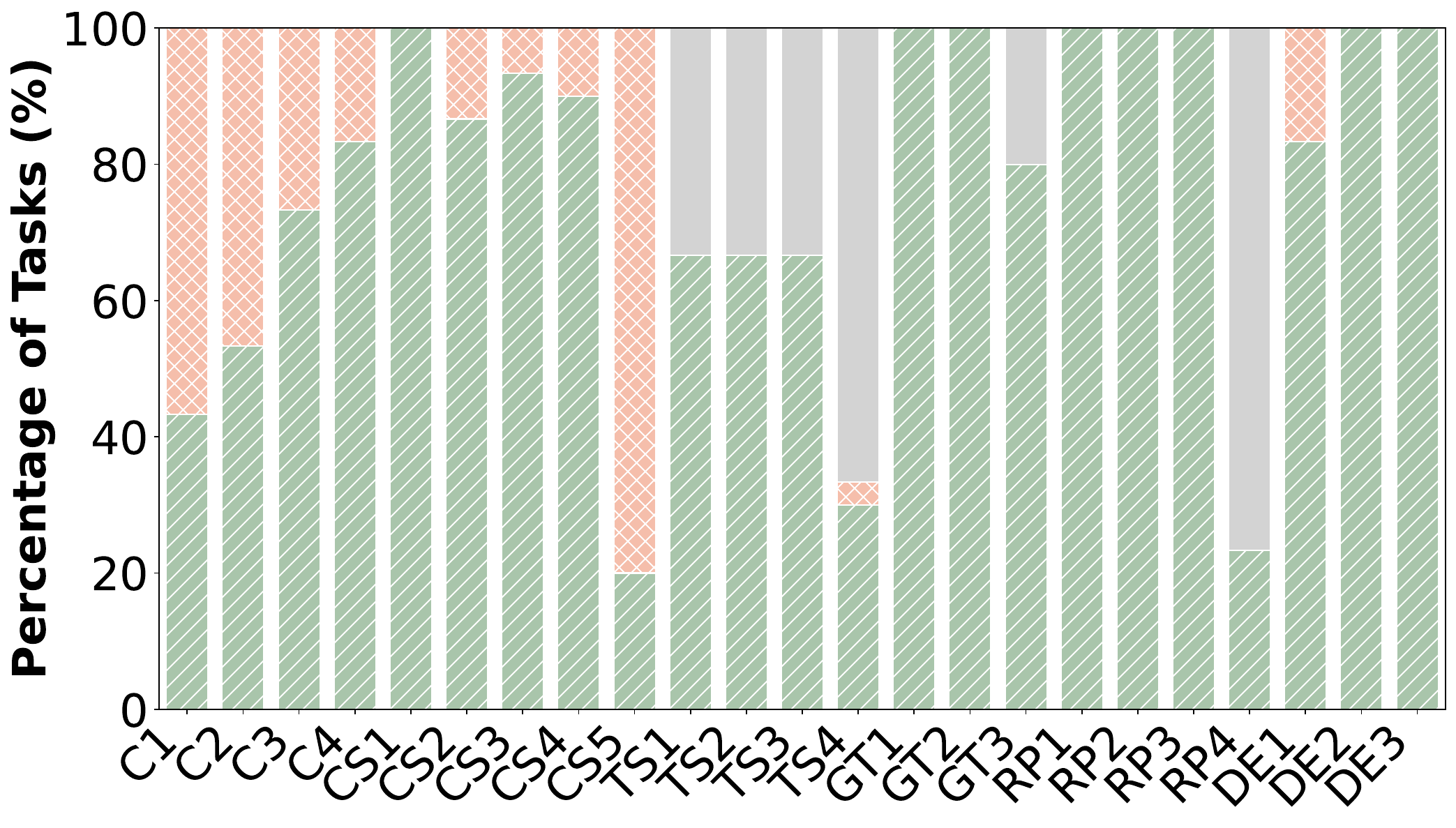}
        \centering
        \includegraphics[width=0.7\textwidth]{plots/eval_proportion_majority_legend.pdf}
        \caption{Pass/Fail rates with majority voting}
        \label{subfig:majority_eval}
    \end{subfigure}\hfill
    \begin{subfigure}[t]{0.48\textwidth}
        \includegraphics[width=\textwidth]{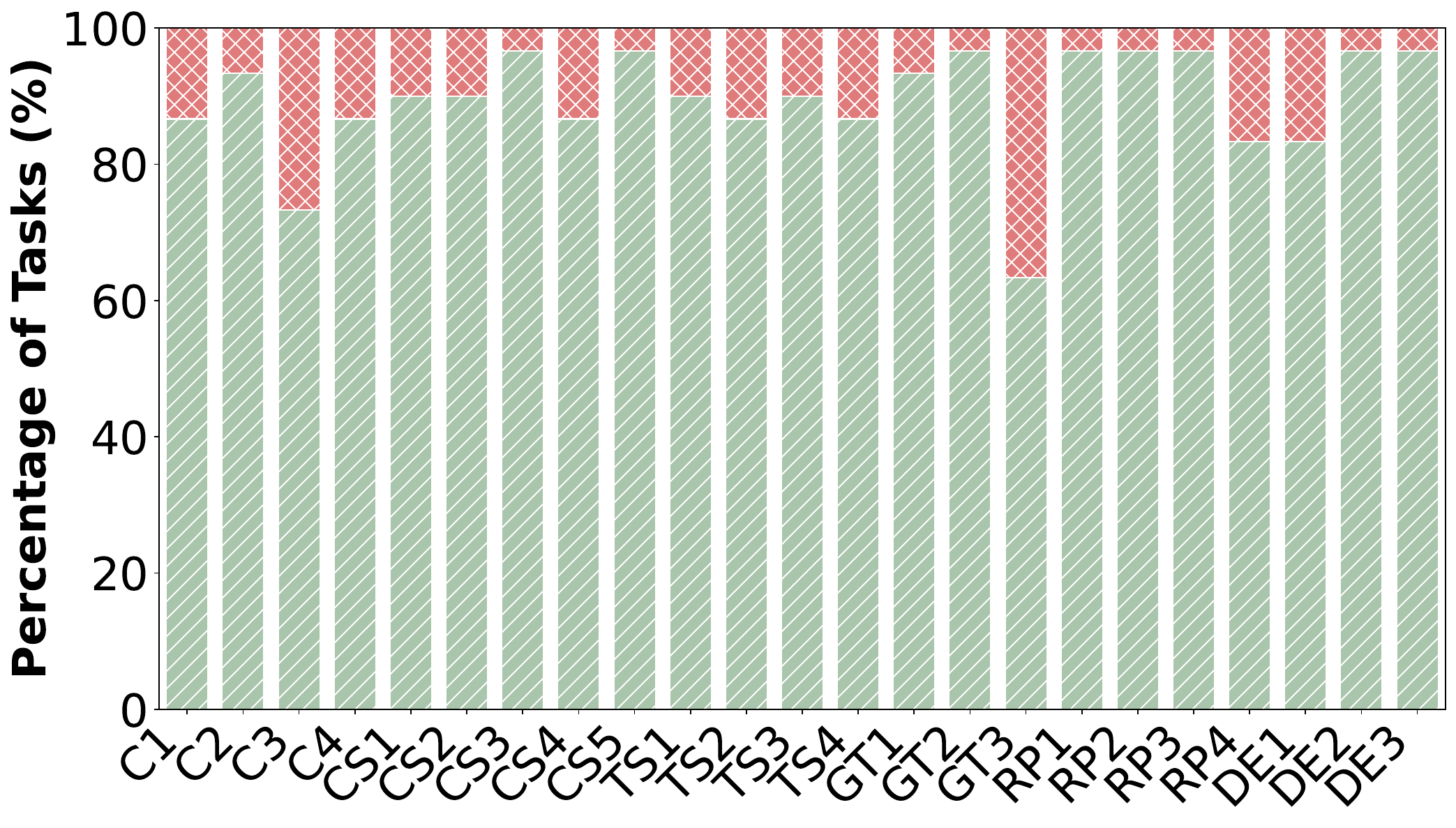}
        \centering
        \includegraphics[width=0.75\textwidth]{plots/majority_vs_human_legend.pdf}
        \caption{Agreement with human judges}
        \label{subfig:majority_human}
    \end{subfigure}

    \caption{Evaluation results using majority voting. (a) Pass/Fail rates per metric when using majority vote (task fails if at least 2 of 3 runs fail). (b) Agreement between majority vote results and human evaluations, achieving 89.4\% overall agreement.}
    \label{fig:majority_vote}
\end{figure}

To show more determined failures, we also apply majority voting to failures, where a task is marked as FAIL when there are two or more runs return FAIL. Issues identified through majority voting represent more consistent failures that persist across multiple evaluation runs.
As shown in Figure~\ref{fig:majority_vote}, majority voting yields an overall higher pass rate. The most common failures remain in \texttt{CS5} (statistical significance) with 80\% failure rate, followed by failures in execution quality. Agreement with human judges improves to 89.4\% when using majority voting.


\end{document}